\documentclass[12pt]{article}
\usepackage{amsmath,epsfig,amssymb,amsfonts,amsthm,epsfig,verbatim, enumitem}
\usepackage{color,graphicx,lscape,longtable,mathabx,natbib,float}
\usepackage{multirow}
\usepackage{setspace}

\usepackage[utf8]{inputenc}

\newtheorem{Theorem}{Theorem}[section]

\def\bse{\begin{eqnarray*}}
\def\ese{\end{eqnarray*}}
\def\be{\begin{eqnarray}}
\def\ee{\end{eqnarray}}
\def\bsq{\begin{equation*}}
\def\esq{\end{equation*}}
\def\bq{\begin{equation}}
\def\eq{\end{equation}}

\def\cov{\hbox{cov}}

\def\wh{\widehat}
\def\wt{\widetilde}

\def\n{\nonumber}

\def\vecl{\mbox{vecl}}

\def\cov{\mbox{cov}}

\def\vec{\mbox{vec}}

\def\sumi{\sum_{i=1}^n}
\def\sumj{\sum_{j=1}^n}
\def\trans{^{\rm T}}

\def\ba{\boldsymbol\alpha}

\def\bb{\boldsymbol\beta}
\def\bg{\boldsymbol\gamma}

\def\0{{\bf 0}}
\def\1{{\bf 1}}
\def\A{{\bf A}}
\def\a{{\bf a}}
\def\m{{\bf m}}
\def\B{{\bf B}}
\def\C{{\bf C}}
\def\c{{\bf c}}
\def\D{{\bf D}}

\def\b{{\bf b}}

\def\X{{\bf X}}
\def\x{{\bf x}}

\def\z{{\bf z}}

\def\bq{\begin{equation}}
\def\eq{\end{equation}}
\def\pr{\hbox{pr}}
\def\wh{\widehat}
\def\wt{\widetilde}
\def\trans{^{\rm T}}

\def\squarebox#1{\hbox to #1{\hfill\vbox to #1{\vfill}}}

\def\boeta{{\boldsymbol \eta}}

\def\vec{\mathrm{vec}}

\def\cov{\hbox{cov}}

\def\bse{\begin{eqnarray*}}
\def\ese{\end{eqnarray*}}
\def\be{\begin{eqnarray}}
\def\ee{\end{eqnarray}}
\def\bsq{\begin{equation*}}
\def\esq{\end{equation*}}
\def\bq{\begin{equation}}
\def\eq{\end{equation}}
\def\pr{\hbox{pr}}
\def\wh{\widehat}
\def\wt{\widetilde}

\def\trans{^{\rm T}}

\def\boxit#1{\vbox{\hrule\hbox{\vrule\kern6pt\vbox{\kern6pt#1\kern6pt}\kern6pt\vrule}\hrule}}

\begin{document}
\thispagestyle{empty}
\baselineskip 17pt
\renewcommand {\thepage}{}
\include{titre}
\pagenumbering{arabic}
\begin{center}
{\Large\bf Sufficient dimension reduction for feasible and robust
  estimation of average causal effect} 
\end{center}

\begin{center}
\textsc{By Trinetri Ghosh,  Yanyuan Ma}\\
\textit{Pennsylvania State University \\
University Park, PA 16802, USA \\
tbg5133@psu.edu \ \  yzm63@psu.edu
} \\ 

\textsc{and Xavier de Luna}\\
\textit{Ume\r{a} School of Business, Economics and Statistics at
Ume\r{a} University \\
SE-90187 Ume\r{a}, Sweden}\\
xavier.deluna@umu.se
\end{center}

\textbf{Abstract}: 
When estimating the treatment effect in an observational study, 
we use a semiparametric locally efficient dimension reduction approach
to assess both the treatment assignment mechanism and the average
responses in both treated and nontreated groups. We then integrate all
results through imputation, inverse probability
  weighting and doubly
  robust augmentation estimators. Doubly robust estimators are locally
  efficient while imputation estimators are super-efficient when the
  response models are correct. To take advantage of both procedures,
we introduce a shrinkage
  estimator to automatically 
combine the two, which retains
  the double robustness property while improving on the variance when
  the response model is correct.   
  We demonstrate the
  performance of these estimators through simulated experiments and a real
  dataset concerning the effect of maternal smoking on baby birth weight.


\setcounter{page}{1}

\doublespacing

\section{Introduction}
Dimension reduction is a major methodological issue that must be
tackled in modern observational studies where the interest lies in the
estimation of the causal effect of a non-randomized treatment. 
This is due to the increasing availability of health
and administrative registers, giving access to high-dimensional
pre-treatment information sets which can help identifying causal
effects of interest. This paper introduces and studies estimators of
average causal effect of a binary treatment using semi-parametric sufficient dimension
reduction methods.

Dimension reduction for feasible nonparametric and semiparametric
causal inference has only recently been formalized, with most
contributions focusing on covariate selection, i.e. methods to pick up
which covariates are actual confounders that need to be controlled
for, see, e.g., \cite{Grub:vand:2010, delunaet:11, Farrell:15,
  Shortreed:17}. Dimension reduction must consider nuisance
conditional models; the probability of treatment given the covariates
(propensity score), and models for the two potential responses
(i.e. responses under two possible levels of a binary treatment) given
the covariates \citep{delunaet:11}. 
Sufficient dimension reduction \citep{Li1991, LiDuan1991, Cook1998,
  Xiaetal2002, Xia2007, MaZhu2012} constitutes an
alternative to covariate selection which has the advantage that it
can, not only consider covariates in isolation as confounders, but
also accomodate linear combinations of the whole covariate set. Such methods have
only recently attracted attention in semiparametric causal inference,
where \cite{Liuetal2016} considered sufficient dimension reduction for
the estimation of the propensity score only, \cite{Luoetal:17}
considered sufficient dimension reduction for the estimation of the
response models only, while \cite{Maetal:18} considered classical
sufficient dimension in all nuisance models.

In this paper we take a general approach to the estimation of average
causal effect. We first use efficient semiparametric
sufficient dimension reduction methods \citep{mazhu2013, ma2014} in all
nuisance models explaining the potential responses and the treatment
assignment,  and then combine these into classical imputation (IMP) and
inverse probability weighting (IPW) estimators.
 While our semiparametric
sufficient dimension reduction modelling is very flexible, nuisance
models may still be misspecified and thus a double robust estimator
(augmented inverse probability weighting estimator) is also considered
which allows for the misspecification of one of the nuisance
model. The augmented inverse probability weighting (AIPW) estimator is
locally efficient, in the sense that it
reaches efficiency at the true nuisance
models, while the imputation estimator is super-efficient in the sense
that if the true response model is known then this
knowledge yields a lower asymptotic efficiency bound than the
AIPW estimator may reach \citep{tancom2006}. We therefore
propose a novel estimator shrinking the imputation and AIPW
estimators towards each other. The shrinkage estimator is also double robust. It is
asymptotically equivalent to the AIPW estimator if
the response model is misspecified, and if all nuisance models are
correctly specified it shrinks  towards the imputation estimator 
which is more efficient than AIPW in this case.
In general, it generates an estimator that has no larger variability than both AIPW and IMP.


\section{Model and Dimension Reduction}\label{sec:prep}

Let $Y_T$ be the treatment response under treatment $T$, where $T=1$ if
the treatment of interest is applied and $T=0$ if some 
alternative treatment, for example, placebo or no treatment is applied.
Let $\X\in{\cal R}^p$ be the set of pre-treatment covariates. We
observe a random sample $\{\X_i, T_i, Y_{1i}T_i+Y_{0i}(1-T_i) \}$, for
$i=1, \dots, n$.  In particular, $Y_{ti}$ is observed only for unit
$i$ such that $T_i=t$, and are therefore called potential responses.
Our goal is to 
estimate the average causal effect of the treatment, here $D=E(Y_1-Y_0)$. We
assume $0<\pr(T=1\mid Y_0,Y_1,\X)=\pr(T=1\mid \X)<1$ throughout.
This assumption is often called strong ignorability of the treatment
assignment, and yields identification of the parameter $D$ under the
above sampling scheme \citep[e.g.,][]{RR1983}.

We now describe flexible dimension reduction structures that will be
combined into different semiparametric estimators for $D$.
First, the treatment assignment probability, also called propensity
score in the literature, can be modelled as 
\be
\pr(T=1\mid
\X=\x)=e^{\eta(\ba\trans\x)}/\{1+e^{\eta(\ba\trans\x)}\},\label{eq:treatment}
\ee
where $\eta(\cdot)$ is an unknown function, smooth and bounded from
both above and below to guarantee the propensity is strictly in
$(0,1)$, and $\ba$ is an unknown index 
vector or matrix with dimension $p\times d_\alpha$, $p>d_\alpha$. 

Further, we model $Y_1$ given $\X=\x$ using a flexible
dimension reduction model
\be
Y_1=m_1(\bb_1\trans\x)+\epsilon_1. \label{eq:model1}
\ee
where $E(\epsilon_1\mid\x)=0$.
Similarly, we model $Y_0$ given $\X=\x$ via
\be
Y_0=m_0(\bb_0\trans\x)+\epsilon_0, \label{eq:model0}
\ee
where
$E(\epsilon_0\mid\x)=0$. Here, $m_1(\cdot), m_0(\cdot)$ are
unknown functions, and $\bb_1, \bb_0$ are unknown index vectors or
matrices with dimension $p\times d_1$ and $p\times d_0$ respectively,
for $p>d_1, p>d_0$.

The models (\ref{eq:treatment}), (\ref{eq:model1}) and (\ref{eq:model0}) 
 separately describe the probability of receiving
treatment and the mean potential responses without imposing any
relation between these models.
Hence, based on each of the three models, 
we can estimate the corresponding unknown parameters
and unknown functions involved in the models 
separately using a random sample.
We can then combine these estimators in various ways to estimate the
treatment effect $D=E(Y_1-Y_0)$.

\subsection{Estimation of Response Models}\label{subsec:imp1}

We first consider (\ref{eq:model1}). Because of the ignorability of
the treatment assignment assumption, the subset of the sample that are
treated indeed form a
random sample to fit model (\ref{eq:model1}). Thus, we can directly implement
the semiparametric method of \cite{ma2014} for the estimation of both
$\bb_1$ and $m_1(\cdot)$, based on the subset of the data with $T_i=1$.
For identifiability reason, we adopt the parameterization of
\cite{ma2014} and fix the upper $d_1 \times d_1$ submatrix of $\bb_1$
as the identity matrix and leave the lower  $(p-d_1)\times d_1$ 
submatrix arbitrary. The locally efficient estimator of $\bb_1$ is thus
obtained from solving
\be \label{eq:b1}
\sumi t_i\{y_{1i}-\wh m_1(\bb_1\trans\x_i,\bb_1)\}\wh
\m_1'(\bb_1\trans\x_i,\bb_1) \otimes\{\x_{Li}-\wh E(\X_{Li}\mid\bb_1\trans\x_i)\}=\0,
\ee
where the Nadaraya-Watson kernel estimator is used to obtain $\wh E(\X_L
\mid \bb_1\trans\x)$ and the local linear estimator is used to obtain $\wh m_1(\bb_1\trans\x,\bb_1)$ and  
$\wh \m_1'(\bb_1\trans\x,\bb_1)$, where $\X_L$ represents the subvector of $\X$
formed by the lower $p-d_1$ components.  Specifically, in (\ref{eq:b1}),
\bse
\wh  E(\X_{L}\mid\bb_1\trans\x)
=\frac{\sumi \x_{Li}K_h(\bb_1\trans\x_i-\bb_1\trans\x)}
{\sumi K_h(\bb_1\trans\x_i-\bb_1\trans\x)},
\ese
and  $\wh m_1(\bb_1\trans\x, \bb_1)=c_0, \wh\m_1'(\bb_1\trans\x,
\bb_1)=\c_1$ are the solution to 
\be\label{eq:m1.eq}
\min_{c_0, \c_1}\sumi t_i\{y_{1i}-c_0-\c_1\trans(\bb_1\trans\x_i-\bb_1\trans\x)\}^2
K_h(\bb_1\trans\x_i-\bb_1\trans\x).
\ee
It is easy to verify that the minimizer of (\ref{eq:m1.eq}) has the
explicit form
\be
\wh m_1(\bb_1\trans\x, \bb_1)&=& A_{11} + \A_{13}\trans
(\A_{14}-\A_{13}\A_{13}\trans)^{-1} \A_{13} A_{11},\label{eq:m1} \\
\wh \m_1'(\bb_1\trans\x, \bb_1)&=& (\A_{14}-\A_{13}\A_{13}\trans)^{-1}(\A_{12}-\A_{13}A_{11}),\n
\ee
where
\begin{align*}
& A_{11} = \frac{\sumi t_i y_{1i} K_h(\bb_1\trans\x_i -\bb_1\trans\x) }{\sumi t_i K_h(\bb_1\trans\x_i-\bb_1\trans\x) }, 
& & \A_{12} = \frac{\sumi t_i y_{1i} (\bb_1\trans\x_i - \bb_1\trans\x) K_h(\bb_1\trans\x_i - \bb_1\trans\x) }{ \sumi t_i K_h(\bb_1\trans\x_i-\bb_1\trans\x)}, \\
& \A_{13} = \frac{\sumi t_i (\bb_1\trans\x_i - \bb_1\trans\x)
  K_h(\bb_1\trans\x_i - \bb_1\trans\x)}{\sumi t_i K_h(\bb_1\trans\x_i-\bb_1\trans\x)}, & &
\A_{14} = \frac{\sumi t_i (\bb_1\trans\x_i -\bb_1\trans\x)^{\otimes2} K_h(\bb_1\trans\x_i - \bb_1\trans\x)}{\sumi t_i K_h(\bb_1\trans\x_i-\bb_1\trans\x)},
\end{align*}
and $\a^{\otimes2}=\a\a\trans$ throughout the text.
Note that the above description is a typical profiling estimation
procedure for $\bb_1$. Once we obtain $\wh\bb_1$, we then
estimate $m_1$ using $\wh m_1(\wh\bb_1\trans\x, \wh\bb_1)$ given in (\ref{eq:m1}).

Theorem 1 of \cite{ma2014} established the property of the above
estimator. Specifically, the estimator $\wh\bb_1$ satisfies
\be\label{eq:useful1}
&&\sqrt{n_1}\vecl(\wh\bb_1-\bb_1)\\
&=&-\B_1n_1^{-1/2}\sumi t_i\{y_{1i}-m_1(\bb_1\trans\x_i)\}
\vec[\m_1'(\bb_1\trans\x_i) 
\otimes\{\x_{Li}-E(\X_{Li}\mid\bb_1\trans\x_i)\}]+o_p(1),\n
\ee
where $n_1=\sumi T_i$, $\vecl(\bb_1)$ is the vector formed by the lower
$(p-d_1)\times d_1$ submatrix of $\bb_1$, and 
\be\label{eq:matB1}
\B_1&\equiv&\left\{E\left(\frac{\partial\vec[T_i\{Y_{1i}-m_1(\bb_1\trans\X_i)\}
\m_1'(\bb_1\trans\X_i) \otimes\{\X_{Li}-
E(\X_{Li}\mid\bb_1\trans\X_i)\}]}{\partial\vecl(\bb_1)\trans}\right)\right\}^{-1}.
\ee

Similar analysis can be used to estimate $\bb_0$ and
$m_0$,  using the subset of the dataset corresponding to
$T_i=0$. Then implementing Theorem 1 from \cite{ma2014}, the
asymptotic behavior of the efficient estimator $\wh\bb_0$
is given by 
\be\label{eq:useful2}
&&\sqrt{n_0}\vecl(\wh\bb_0-\bb_0) \\
&=&-\B_0
 n_0^{-1/2}\sumi (1-t_i)\{y_{0i}-m_0(\bb_0\trans\x_i)\}
\vec[\m_0'(\bb_0\trans\x_i) 
\otimes\{\x_{Li}-E(\X_{Li}\mid\bb_0\trans\x_i)\}]+o_p(1),\n
\ee
where $n_0=n-n_1$, and
\be\label{eq:matB0}
\B_0&\equiv&\left\{E\left(\frac{\partial\vec[(1-T_i)\{Y_{0i}-m_0(\bb_0\trans\X_i)\}
\m_0'(\bb_0\trans\X_i) \otimes\{\X_{Li}-
E(\X_{Li}\mid\bb_0\trans\X_i)\}]}{\partial\vecl(\bb_0)\trans}\right)\right\}^{-1}.
\ee
When the mean function models are correct, the meaning of $\bb_1$,
$\bb_0$, $m_1$ and $m_0$ is easy to understand. When the models
are incorrect, as we shall allow in the sequel, we can understand
$\bb_1$, $\bb_0$, $m_1$ and $m_0$ as
quantities that satisfy 
\bse
E[ T\{Y_{1}-m_1(\bb_1\trans\X,\bb_1)\}
\m_1'(\bb_1\trans\X,\bb_1)
\otimes\{\X_{L}-E(\X_{L}\mid\bb_1\trans\X)\}]=\0,\\
E[ (1-T)\{Y_0-m_0(\bb_0\trans\X,\bb_0)\}
\m_0'(\bb_0\trans\X,\bb_0)
\otimes\{\X_L-E(\X_L\mid\bb_0\trans\X)\}]=\0,
\ese
where
$
m_1(\bb_1\trans\x) = E(Y_1 \mid \bb_1\trans\x) \neq E(Y_1\mid \x)$, and
$m_0(\bb_0\trans\x) = E(Y_0 \mid \bb_0\trans\x) \neq E(Y_0\mid \x).
$

\subsection{Estimation of Propensity Score Model}\label{subsec:ipw}

The estimation of $\ba,\eta$ was also studied in the literature
\citep{Liuetal2016,mazhu2013}, hence we directly write out the five step 
algorithm here for completeness of the content and clarity. 

\begin{description}

\item[Step 1.] Form the Nadaraya-Watson estimator of
  $E(\X_i\mid\ba\trans\x_i)$ to obtain
  $\wh{E}(\X_i\mid\ba\trans\x_i)$.

\item[Step 2.] Solve 
$
\sumi
\vecl ( \{\x_i-\wh{E}(\X_i\mid\ba\trans\x_i)\}[t_i-1+1/\{1+\exp(\1_d\trans\ba\trans\x_i)\}]\1_d\trans)=\0
$ 
to obtain a consistent initial estimator  $\wt\ba$.

\item[Step 3.] Obtain the local linear estimators of 
  $\eta(\z,\ba)$ and its first derivative 
  $\eta'(\z,\ba)$ by solving
  \be\label{eq:eta.eq}
\sumi\left[t_i-\frac{\exp\{b_{0}+\b_{1}\trans(\ba\trans\x_i-\z)\}}{1+\exp\{b_{0}+\b_{1}\trans(\ba\trans\x_i-\z)\}}\right] K_{h}(\ba\trans\x_i-\z)&=&0 \n \\
\sumi\left[t_i-\frac{\exp\{b_{0}+\b_{1}\trans(\ba\trans\x_i-\z)\}}{1+\exp\{b_{0}+\b_{1}\trans(\ba\trans\x_i-\z)\}}\right](\ba\trans\x_i-\z)\label{eq:eta'}
 K_{h}(\ba\trans\x_i-\z)&=&\0,
\ee
for $b_0,\b_1$ at $\z=\ba\trans\x_1, \dots, \ba\trans\x_n$.
 Write the resulting
  estimator as $\wh{\eta}(\ba\trans\x_i,\ba)$ and $\wh{\boeta}'(\ba\trans\x_i,\ba)$.

\item[Step 4] Insert $\wh{\eta}(\cdot,\ba)$, $\wh{\boeta}'(\cdot,\ba)$ and
  $\wh{E}(\cdot)$ into the estimating
  equation 
\bse
\sumi
 \{\x_{Li}-\wh E(\X_{Li}\mid\ba\trans\x_i)\}\left[t_i-\frac{\exp\{\wh\eta(\ba\trans\x_i)\}}{1+\exp\{\wh\eta(\ba\trans\x_i)\}}\right]\wh\boeta'(\ba\trans\x_i)\trans
=\0
\ese
and solve it to obtain the efficient estimator $\wh{\ba}$, using starting value $\wt\ba$.

\item[Step 5]
Repeat Step 3 at $\ba=\wh\ba$ to obtain the final estimator of $\eta(\cdot)$.
	
\end{description}

We will then form 
$\wh\pr(T=1\mid\X=\x)=\exp\{\wh\eta(\wh\ba\trans\x)\}/[1+\exp\{\wh\eta(\wh\ba\trans\x)\}]$
and use it in the final calculation of the average causal effect.
Let us write
\bse
p_i=\frac{\exp\{\eta(\ba\trans\x_i)\}}{1+\exp\{\eta(\ba\trans\x_i)\}},
P_i=\frac{\exp\{\eta(\ba\trans\X_i)\}}{1+\exp\{\eta(\ba\trans\X_i)\}},
\ese
and define
\be\label{eq:matB}
\B&\equiv&\left\{
 E\left(\frac{\partial}{\partial\vecl(\ba)\trans}
\vec\left[
  \{\X_{Li}-E(\X_{Li}\mid\ba\trans\X_i)\}(T_i-P_i)\boeta'(\ba\trans\X_i)\trans
 \right]\right)
 \right\}^{-1}.
\ee
Then using Lemma 2 from \cite{Liuetal2016}, we have 
\be \label{eq:useful3}
\sqrt{n}\vecl(\wh\ba-\ba)
=
 -\B n^{-1/2}\sumi
 (t_i-p_i)
 \vec[\{\x_{Li}-E(\X_{Li}\mid\ba\trans\x_i)\}
\boeta'(\ba\trans\x_i)\trans]+o_p(1).
\ee
When the propensity score model is correct, the meaning of $\ba$ and $\eta$ is
clear. When the model is incorrect, as we shall allow in the sequel,
$\ba$ and $\eta$ are the quantities
that satisfy
\bse
E[
 \{\X_L-E(\X_L\mid\ba\trans\X)\}\left[T-\frac{\exp\{\eta(\ba\trans\X)\}}{1+\exp\{\eta(\ba\trans\X)\}}\right]\boeta'(\ba\trans\X)\trans]
=\0
\ese
where
$[1+\exp\{\eta(-\ba\trans\x)\}]^{-1}=E(T\mid\ba\trans\x)\ne E(T\mid\x).$

\section{Average Causal Effect: Estimators and Properties}\label{sec:dy}

We  are now ready to propose several estimators for estimating the average
treatment effect, based on the semiparametric modeling and estimators
described in Section \ref{sec:prep}. These propositions all take advantage of existing
methods in missing at random problems, including imputation and
weighting, hence they inherit the 
properties expected. 
We also introduce a novel shrinkage estimator combining imputation and
weighting, with an optimal property. Let $y_i=t_iy_{1i}+(1-t_i)y_{0i}$ be the observed
response value.

\subsection{Imputation Estimators}
First we consider estimating the average causal
effect using an imputation approach,
first proposed in the context of missing data  \citep{rubin1978}.
The imputation approach we take here is semiparametric in a spirit  similar to
 the nonparametric imputation \citep{wgm2012}. Specifically, we construct
\bse 
\wh E(Y_1)
&=&n^{-1}\sumi \left\{t_iy_i+(1-t_i) 
\wh m_1(\wh\bb_1\trans\x_i)\right\},\\
\wh E(Y_0)
&=&n^{-1}\sumi \left\{(1-t_i)y_i+t_i
\wh m_0(\wh\bb_0\trans\x_i)\right\},
\ese 
and then form the imputation estimator IMP as $\wh D_{\rm IMP}=\wh E(Y_1) - \wh E(Y_0)$.

We further consider an alternative imputation estimator which uses the
model predicted values while ignoring the observed
responses even when they are available. Specifically, we still form
$\wh D_{\rm IMP2}\equiv\wh E(Y_1)-\wh E(Y_0)$ for the 
treatment effect, while using
\bse 
\wh E(Y_1)
=n^{-1}\sumi  
\wh m_1(\wh\bb_1\trans\x_i),\ \ \ 
\wh E(Y_0)
=n^{-1}\sumi 
\wh m_0(\wh\bb_0\trans\x_i),
\ese
 to obtain the imputation estimator IMP2.
The latter is sometimes named outcome regression estimator, see for
example \cite{tancom2006}.

\subsection{(Augmented) Inverse Probability Weighting Estimators}

\cite{rrzhao1994} proposed a class of semiparametric estimators based
on inverse probability weighted (IPW) estimating equations, borrowing the
idea 
of \cite{ht1952} in the survey sampling literature. Later
\cite{Liuetal2016} implemented the IPW estimator with
semiparametric modeling to assess the
propensity score function. Following this procedure, the IPW estimator consists
in constructing
\bse 
\wh E(Y_1)
&=&n^{-1}\sumi
\frac{t_iy_i [1+\exp\{\wh\eta(\wh\ba\trans\x_i)\}]}{
  {\exp\{\wh\eta(\wh\ba\trans\x_i)\}}},\\
\wh E(Y_0)
&=&n^{-1}\sumi
(1-t_i)y_i [1+\exp\{\wh\eta(\wh\ba\trans\x_i)\}],
\ese 
and then form the estimate of the average causal effect $\wh D_{\rm
  IPW}\equiv\wh E(Y_1)-\wh E(Y_0)$. 


If at least one of the mean function models, $m_1(\cdot)$ and
$m_0(\cdot)$, is
incorrectly specified, the IMP and IMP2 estimators will be
inconsistent. Similarly if $\eta(\cdot)$ is incorrectly specified IPW
is not consistent. Because of this, we have used more flexible 
semiparametric dimension reduction models instead of fully
parametric models. However, this lowers, but does not completely eliminate, the
chance of model misspecification. 
 Thus, protection from either misspecification
via the doubly robust estimator \citep{rrzhao1994} is still
desired. This leads to the
 augmented inverse probability weighting estimator (AIPW), which has the
 property of consistency when either
the mean models are correctly specified or the propensity score model
is correctly specified. The estimate of average causal effect is still
$\wh D_{\rm AIPW}\equiv\wh E(Y_1)-\wh E(Y_0)$, where now
\bse 
\wh E(Y_1)
&=&n^{-1}\sumi \left\{
\frac{t_iy_i [1+\exp\{\wh\eta(\wh\ba\trans\x_i)\}]}{
  {\exp\{\wh\eta(\wh\ba\trans\x_i)\}}}
+\left(1-\frac{t_i [1+\exp\{\wh\eta(\wh\ba\trans\x_i)\}]}{\exp\{\wh\eta(\wh\ba\trans\x_i)\}}\right)
\wh m_1(\wh\bb_1\trans\x_i)\right\}\\
\wh E(Y_0)
&=&n^{-1}\sumi \left\{
(1-t_i)y_i [1+\exp\{\wh\eta(\wh\ba\trans\x_i)\}]
+\left(1-(1-t_i) [1+\exp\{\wh\eta(\wh\ba\trans\x_i)\}]
\right)
\wh m_0(\wh\bb_0\trans\x_i)\right\}.
\ese 


An improved version of the AIPW estimator was proposed in
\cite{rrzhao1995}, which provides extra protection against deteriorated
estimation variability. Based on this idea, \cite{tan2006} later
developed a nonparametric likelihood estimator.  
Adopting this idea in the treatment
effect estimation framework, we construct the estimator
\bse 
\wh E(Y_1)
&=&n^{-1}\sumi \left\{
\frac{t_iy_i [1+\exp\{\wh\eta(\wh\ba\trans\x_i)\}]}{
  {\exp\{\wh\eta(\wh\ba\trans\x_i)\}}}
+\wh\gamma_1\left(1-\frac{t_i [1+\exp\{\wh\eta(\wh\ba\trans\x_i)\}]}{\exp\{\wh\eta(\wh\ba\trans\x_i)\}}\right)
\wh m_1(\wh\bb_1\trans\x_i)\right\}\\
\wh E(Y_0)
&=&n^{-1}\sumi \left\{
(1-t_i)y_i [1+\exp\{\wh\eta(\wh\ba\trans\x_i)\}]
+\wh\gamma_0\left(1-(1-t_i) [1+\exp\{\wh\eta(\wh\ba\trans\x_i)\}]\right)
\wh m_0(\wh\bb_0\trans\x_i)\right\},
\ese 
and  estimate the average causal effect by $\wh D_{\rm IAIPW}\equiv\wh E(Y_1)-\wh E(Y_0)$.
Here
\bse
\wh\gamma_1&=&
\cov \left\{m_1(\wh\bb_1\trans\x_i)
\frac{t_i
 [1+\exp\{\wh\eta(\wh\ba\trans\x_i)\}]}{
  {\exp\{\wh\eta(\wh\ba\trans\x_i)\}}},\left(1-\frac{t_i
      [1+\exp\{\wh\eta(\wh\ba\trans\x_i)\}]}{\exp\{\wh\eta(\wh\ba\trans\x_i)\}}\right)\wh
  m_1(\wh\bb_1\trans\x_i)\right\}^{-1}\\
&&\times\cov \left\{
\frac{t_iy_i [1+\exp\{\wh\eta(\wh\ba\trans\x_i)\}]}{
  {\exp\{\wh\eta(\wh\ba\trans\x_i)\}}},\left(1-\frac{t_i
      [1+\exp\{\wh\eta(\wh\ba\trans\x_i)\}]}{\exp\{\wh\eta(\wh\ba\trans\x_i)\}}\right)\wh
  m_1(\wh\bb_1\trans\x_i)\right\},\\
\wh\gamma_0&=&
\cov\left\{(1-t_i) \wh m_0(\wh\bb_0\trans\x_i) [1+\exp\{\wh\eta(\wh\ba\trans\x_i)\}],\left(1-(1-t_i) [1+\exp\{\wh\eta(\wh\ba\trans\x_i)\}]\right)
\wh m_0(\wh\bb_0\trans\x_i)\right\}
^{-1}\\
&&\times\cov\left\{(1-t_i)y_i [1+\exp\{\wh\eta(\wh\ba\trans\x_i)\}],
\left(1-(1-t_i) [1+\exp\{\wh\eta(\wh\ba\trans\x_i)\}]\right)
\wh m_0(\wh\bb_0\trans\x_i)\right\}.
\ese

\subsection{The Shrinkage  Estimator}

The ideas of imputation and weighting are quite different and each has
its own advantage and drawback. For example, 
when the treatment mean models
$m_1(\bb_1\trans\X), m_0(\bb_0\trans\x)$ are correct, regardless if the
propensity score model is correct or not,
both  IMP and AIPW  are consistent but it is unclear which estimator
is more efficient. However, when 
the treatment mean models
$m_1(\bb_1\trans\X), m_0(\bb_0\trans\x)$ are not both correct, 
AIPW is still consistent as long as the propensity score model is
correct, while IMP methods will be inconsistent. Of course, if 
both the mean models and the propensity models are incorrect, then
neither methods will provide consistent estimation. In applications, we typically do not know which scenario we are in, hence it is 
hard to determine whether IMP methods or AIPW methods 
are beneficial to use. Because of this situation, in order to take
advantage of both methods, we use the idea of
shrinkage estimator 
\citep{muk_chat2008} to
construct a weighted average 
between IMP and AIPW.

The general observation is that if IMP is consistent, then AIPW is
also automatically consistent, but 
not the other way round. However, it is not generally clear which
estimator is more efficient.
We construct the following shrinkage estimator:
Let
$\sqrt{n}(\wh D_{\rm AIPW}-D_{\rm AIPW})\to N(0, v_{\rm AIPW})$ in distribution,
$\sqrt{n}(\wh D_{\rm IMP}-D_{\rm IMP})\to N(0, v_{\rm IMP})$ in
distribution, 
and let 
$\cov\{\sqrt{n}(\wh D_{\rm AIPW}-D_{\rm AIPW}), \sqrt{n}(\wh
D_{\rm IMP}-D_{\rm IMP})\}\to v_{\rm AI}$.
We form
\bse
w=\frac{(\wh D_{\rm AIPW}-\wh D_{\rm IMP})^2+(v_{\rm IMP}-v_{\rm AI})/\sqrt{n}}
{(\wh D_{\rm AIPW}-\wh D_{\rm IMP})^2+(v_{\rm IMP}+v_{\rm AIPW}-2v_{\rm AI})/\sqrt{n}},
\ese
and form the shrinkage estimator
\bse
\wh D=w\wh D_{\rm AIPW}+(1-w)\wh D_{\rm IMP},
\ese
where we replace $v_{\rm AIPW}, v_{\rm IMP}, v_{\rm AI}$ with their estimated
version. We can see that this construction has the property that 
 when IMP is inconsistent while AIPW is consistent, $w\to 1$
and we essentially obtain AIPW, i.e. the shrinkage estimator is double robust. On the other hand, when both
estimators are consistent, 
\bse
w\to \left\{w_0\equiv\frac{v_{\rm IMP}-v_{\rm AI}}
{v_{\rm IMP}+v_{\rm AIPW}-2v_{\rm AI}}\right\},
\ese
in probability, which yields the optimal combination of the two estimators in terms of
the final estimation variability. Of course when both estimators are
inconsistent, the weighted average is still inconsistent.

To construct the shrinkage estimator described above,  we derived the
asymptotic 
variances and covariances of the estimators in Section \ref{sec:asym}.
Note that one may also choose to shrink IMP2 and AIPW or any of the
two versions of the imputation estimator with the improved AIPW in a similar fashion.

\subsection{Asymptotic properties of the treatment effect estimators}\label{sec:asym} 

In this section, we discuss the asymptotic properties of the average
treatment effect estimators introduced. 
These properties are developed under the following conditions:
\begin{enumerate}[label=C\arabic*]
\item\label{ass:kernel}
The univariate $m$th order kernel function $K(\cdot)$ is symmetric, 
  Lipschitz continuous on its support $[-1,1]$, which satisfies 
  \bse 
  \int K(u) du= 1,  \ \int u^iK(u) du= 0, 1\le i\le m-1,   \ 0\ne \int u^m  K(u)du<\infty. 
  \ese

\item\label{ass:bandwidth}
The bandwidths satisfy 
$nh^{2m}\to 0$,
$nh^{2d}\to\infty$.

  \item\label{ass:density}
The probability density functions
  of $\bb_1\trans\x$, $\bb_0\trans\x$  and $\ba\trans\x$, denoted 
  $f\left(\bb\trans\x\right)$,
  $f\left(\ba\trans\x\right)$ and 
$f\left(\ba\trans\x\right)$ with an abuse of notation, 
   are bounded away from 0 and $\infty$.
\end{enumerate}

Let the true average causal effect be $D = E(Y_1-Y_0)$. Then we have the following results.

\begin{Theorem}\label{th:asimp}
Under the regularity conditions \ref{ass:kernel}-\ref{ass:density}, when $n\rightarrow \infty$, the IMP 
estimator $\wh D_{\rm IMP}$ satisfies 
$
\sqrt{n}(\wh D_{\rm IMP} - D) \overset{d}{\rightarrow} {\rm N}(0,v_{\rm IMP}),
$
where combining the results regarding $\wh E(Y_1)$ and $\wh E(Y_0)$ in
Appendix \ref{sec:proofIMP}, we
get
\be\label{eq:imp}
v_{\rm IMP}&=&E(\sqrt{n}[\{\wh E(Y_1)-E(Y_1)\}-\{\wh E(Y_0)-E(Y_0)\}])^2 \\
&=&  E \big(
 \left\{
m_1(\bb_1\trans\x_i) 
- m_0(\bb_0\trans\x_i) -E(Y_1) +E(Y_0)\right\}\n\\
&&+E[1+\exp\{-\eta(\ba\trans\X_i)\}\mid\bb_1\trans\x_i]t_i\{y_{1i}-m_1(\bb_1\trans\x_i)\}\n\\
&&-E[1+\exp\{\eta(\ba\trans\X_i)\}\mid\bb_0\trans\x_i](1-t_i)\{y_{0i}-m_0(\bb_0\trans\x_i)\}\n\\
&&- E[(1-P_i)\vec\{\X_{Li}\m_1'(\bb_1\trans\X_i)\trans\}]\trans
\B_1\n\\
&&\times t_i\{y_{1i}-m_1(\bb_1\trans\x_i)\}
\vec[\m_1'(\bb_1\trans\x_i) 
\otimes\{\x_{Li}-E(\X_{Li}\mid\bb_1\trans\x_i)\}]\n\\
&&
+ E[P_i\vec\{\X_{Li}\m_0'(\bb_0\trans\X_i)\trans\}]\trans
\B_0\n\\
&&\times (1-t_i)\{y_{0i}-m_0(\bb_0\trans\x_i)\}
\vec[\m_0'(\bb_0\trans\x_i) 
\otimes\{\x_{Li}-E(\X_{Li}\mid\bb_0\trans\x_i)\}] \big)^2,\n
\ee
where $\B_1$ and $\B_0$ are defined in (\ref{eq:matB1}) and (\ref{eq:matB0}), respectively.
\end{Theorem}

\begin{Theorem}\label{th:asimp2}
Under the regularity conditions \ref{ass:kernel}-\ref{ass:density}, when $n\rightarrow \infty$, the IMP2
estimator $\wh D_{\rm IMP2}$ satisfies 
$
\sqrt{n}(\wh D_{\rm IMP2} - D) \overset{d}{\rightarrow} {\rm N}(0,v_{\rm IMP2}),
$
where combining the results regarding $\wh E(Y_1)$ and $\wh E(Y_0)$ from Appendix \ref{sec:proofIMP2}, we
get
\bse
v_{\rm IMP2}&=&E\big(\sqrt{n}[\{\wh E(Y_1)-E(Y_1)\}-\{\wh E(Y_0)-E(Y_0)\}]\big)^2\\
&=& E \big( \left\{
m_1(\bb_1\trans\x_i) 
- m_0(\bb_0\trans\x_i) -E(Y_1) +E(Y_0)\right\}\\
&&+E[1+\exp\{-\eta(\ba\trans\X_i)\}\mid\bb_1\trans\x_i]t_i\{y_{1i}-m_1(\bb_1\trans\x_i)\}\\
&&-E[1+\exp\{\eta(\ba\trans\X_i)\}\mid\bb_0\trans\x_i](1-t_i)\{y_{0i}-m_0(\bb_0\trans\x_i)\}\\
&&- E[\vec\{\X_{Li}\m_1'(\bb_1\trans\X_i)\trans\}]\trans
\B_1 t_i\{y_{1i}-m_1(\bb_1\trans\x_i)\}
\vec[\m_1'(\bb_1\trans\x_i) 
\otimes\{\x_{Li}-E(\X_{Li}\mid\bb_1\trans\x_i)\}]\\
&&+ E[\vec\{\X_{Li}\m_0'(\bb_0\trans\X_i)\trans\}]\trans
\B_0 \\
&& \times (1-t_i)\{y_{0i}-m_0(\bb_0\trans\x_i)\}
\vec[\m_0'(\bb_0\trans\x_i) 
\otimes\{\x_{Li}-E(\X_{Li}\mid\bb_0\trans\x_i)\}]  \big)^2,
\ese
where $\B_1$ and $\B_0$ are defined in (\ref{eq:matB1}) and (\ref{eq:matB0}), respectively.
\end{Theorem}

\begin{Theorem}\label{th:asipw}
Under the regularity conditions \ref{ass:kernel}-\ref{ass:density}, when $n \rightarrow \infty$, the IPW estimator 
$\wh D_{\rm IPW}$ satisfies
$
\sqrt{n}(\wh D_{\rm IPW} - D) \overset{d}\rightarrow {\rm N}(0,v_{\rm IPW}),
$
where combining the results of $\wh E(Y_1)$ and $\wh E(Y_0)$ in Appendix \ref{sec:proofIPW}, we get
\bse
v_{\rm IPW}&=&E\big(\sqrt{n}[\{\wh E(Y_1)-\wh E(Y_0)\}-\{E(Y_1)-E(Y_0)\}]\big)^2 \\
&=&E \bigg( \left\{\frac{t_iy_{1i}}{p_i}-E(Y_1)-\frac{(1-t_i)y_{0i}}{1-p_i}+E(Y_0)\right\}\\
&&+ \left(1-\frac{t_i}
{p_i}\right)
E\left\{ m_1(\bb_1\trans\X_i)\mid\ba\trans\x_i\right\} 
- \left(\frac{t_i-p_i}
{1-p_i}\right)
E\left\{ m_0(\bb_0\trans\X_i)\mid\ba\trans\x_i\right\} \\
&&+ \left(
E\left[
\frac{m_{1i}(\bb_1\trans\X_i)+\exp\{\eta(\ba\trans\X_i)\} m_{0i}(\bb_0\trans\X_i)}{1+\exp\{\eta(\ba\trans\X_i)\}}
\vec\{\X_{Li}\boeta'(\ba\trans\X_i)\trans\}
\right]\right)\trans
\B \\
&&\times 
(t_i-p_i)
\vec[\{\x_{Li}-E(\X_{Li}\mid\ba\trans\x_i)\}
\boeta'(\ba\trans\x_i)\trans] \bigg)^2,
\ese
where $\B$ is defined in (\ref{eq:matB}).
\end{Theorem}

\begin{Theorem}\label{th:asaipw}
Under the regularity conditions \ref{ass:kernel}-\ref{ass:density}, when $n \rightarrow \infty$, the AIPW estimator $\wh D_{\rm AIPW}$ satisfies
$
\sqrt{n}(\wh D_{\rm AIPW} - D) \overset{d}\rightarrow {\rm N}(0,v_{\rm AIPW}),
$
where $v_{\rm AIPW}$ derived in Appendix \ref{sec:proofAIPW} is 
\be\label{eq:aipw}
v_{\rm AIPW}&=&E\big(\sqrt{n}[\{\wh E(Y_1)-\wh E(Y_0)\}-\{E(Y_1)-E(Y_0)\}]\big)^2\\
&=&
E  \left(\{y_{1i}-m_1(\bb_1\trans\x_i)\}
t_i [1+\exp\{-\eta(\ba\trans\x_i)\}]+m_1(\bb_1\trans\x_i)-E(Y_1)\right.\n\\
&&-\C_1\B_1t_i\{y_{1i}-m_1(\bb_1\trans\x_i)\}
\vec[\m_1'(\bb_1\trans\x_i) 
\otimes\{\x_{Li}-E(\X_{Li}\mid\bb_1\trans\x_i)\}]\n\\
&&+\D_1\B (t_i-p_i)
\vec[\{\x_{Li}-E(\X_{Li}\mid\ba\trans\x_i)\}
\boeta'(\ba\trans\x_i)\trans]\n\\
&&- \{y_{0i}-m_0(\bb_0\trans\x_i)\}
(1-t_i)
[1+\exp\{\eta(\ba\trans\x_i)\}]-m_0(\bb_0\trans\x_i)+E(Y_0)\n\\
&&+\C_0\B_0
 (1-t_i)\{y_{0i}-m_0(\bb_0\trans\x_i)\}
\vec[\m_0'(\bb_0\trans\x_i) 
\otimes\{\x_{Li}-E(\X_{Li}\mid\bb_0\trans\x_i)\}]\n\\
&&\left.+\D_0\B 
(t_i-p_i)
\vec[\{\x_{Li}-E(\X_{Li}\mid\ba\trans\x_i)\}
\boeta'(\ba\trans\x_i)\trans]\right)^2 ,\n
\ee
where
\bse
\C_1&\equiv&
E \left\{\frac{\partial m_1(\bb_1\trans\X_i) }{\partial\vecl(\bb_1)\trans}
(1-T_i
[1+\exp\{-\eta(\ba\trans\X_i)\}])\right\},\\
\D_1&\equiv&E\left[ \{Y_{1i}-m_1(\bb_1\trans\X_i)\} T_i\exp\{-\eta(\ba\trans\X_i)\}\vec\{\X_{Li}\boeta'(\ba\trans\X_i)\trans\}
\right]\\
\C_0&\equiv&
E \left\{\frac{\partial m_0(\bb_0\trans\X_i) }{\partial\vecl(\bb_0)\trans}
(1-(1-T_i)
[1+\exp\{\eta(\ba\trans\X_i)\}])\right\},\\
\D_0&\equiv&E\left[ \{Y_{0i}-m_0(\bb_0\trans\X_i)\} (1-T_i)\exp\{\eta(\ba\trans\X_i)\}\vec\{\X_{Li}\boeta'(\ba\trans\X_i)\trans\}
\right].
\ese
Note that $\C_1$, $\C_0$, $\D_1$ and $\D_0$ will degenerate to zero if
the relevant model is correct. Then  
\be\label{eq:corrmodaipw}
v_{\rm AIPW}&=&E  \Big(\{y_{1i}-m_1(\bb_1\trans\x_i)\}
t_i[1+\exp\{-\eta(\ba\trans\x_i)\}]+m_1(\bb_1\trans\x_i)-E(Y_1) \\ \nonumber
&-& \{y_{0i}-m_0(\bb_0\trans\x_i)\}
(1-t_i)
[1+\exp\{\eta(\ba\trans\x_i)\}]-m_0(\bb_0\trans\x_i)+E(Y_0) \Big)^2.
\ee
\end{Theorem}

Noting that 
$
\left(1-t_i [1+\exp\{-\eta(\ba\trans\x_i)\}]\right)
m_1(\bb_1\trans\x_i)
$
an
$
\left(1-(1-t_i) [1+\exp\{\eta(\ba\trans\x_i)\}]\right)
m_0(\bb_0\trans\x_i)
$
have mean zero, it is straightforward to show that the 
improved AIPW estimator has the same asymptotic expansion as the AIPW
estimator when all three models are correct. Thus, despite their
different finite sample performance, 
the expansion in (\ref{eq:corrmodaipw}) also applies to the improved AIPW
estimator. 
Thus the following result holds.
\begin{Theorem}\label{th:asiaipw}
Under the regularity conditions \ref{ass:kernel}-\ref{ass:density}
and assuming all models are correct, then when $n \rightarrow \infty$,
the improved AIPW estimator  
$\wh D_{\rm IAIPW}$ satisfies
$
\sqrt{n}(\wh D_{\rm IAIPW} - D) \overset{d}{\rightarrow} {\rm N}(0,v_{\rm AIPW}),
$
where $v_{\rm AIPW}$ is here given by (\ref{eq:corrmodaipw}).
\end{Theorem}

Finally, when both estimators $\hat D_{IMP}$ and $\hat D_{AIPW}$ are consistent, we have
\bse
\sqrt{n}(\wh D-D)
&=&\sqrt{n} w_0(\wh D_{AIPW}-D)+\sqrt{n} (1-w_0)(\wh D_{IMP}-D)+o_p(1),
\ese
as was noted above.
\begin{Theorem}\label{th:asshrink}
Under the regularity conditions
\ref{ass:kernel}-\ref{ass:density}, when $\wh D_{\rm AIPW}$ and
$\wh D_{\rm IMP}$ are consistent and $n \rightarrow \infty$, the
shrinkage estimator $\wh D$ satisfies 
$
\sqrt{n}(\wh D - D) \overset{d}{\rightarrow} {\rm N}(0,v_{\rm shrinkage}),
$
where $v_{\rm shrinkage}= w_0^2 v_{\rm AIPW} + (1-w_0)^2 v_{\rm IMP} +
2 w_0 (1-w_0)v_{\rm AI}$, with
\be
v_{\rm AI}&=&E\left\{
\left(\{y_{1i}-m_1(\bb_1\trans\x_i)\}
t_i
[1+\exp\{-\eta(\ba\trans\x_i)\}]+m_1(\bb_1\trans\x_i)-E(Y_1)\right.\right.
\n\\
&&\left.-
\{y_{0i}-m_0(\bb_0\trans\x_i)\}(1-t_i)
[1+\exp\{\eta(\ba\trans\x_i)\}]-m_0(\bb_0\trans\x_i)+E(Y_0)\right)\n\\
&&\times\left(
t_iy_{1i}-(1-t_i)y_{0i}
+(1-t_i)  m_1(\bb_1\trans\x_i) 
-t_i  m_0(\bb_0\trans\x_i) -E(Y_1) +E(Y_0)\right.\n\\
&&+
E[\exp\{-\eta(\ba\trans\X_i)\}\mid\bb_1\trans\x_i]t_i\{y_{1i}-m_1(\bb_1\trans\x_i)\}\n\\
&&-
E[\exp\{\eta(\ba\trans\X_i)\}\mid\bb_0\trans\x_i](1-t_i)\{y_{0i}-m_0(\bb_0\trans\x_i)\}\n\\
&&- E[(1-P_i)
\vec\{\X_{Li}\m_1'(\bb_1\trans\X_i)\trans\}]\trans
\B_1 t_i\{y_{1i}-m_1(\bb_1\trans\x_i)\}\n\\
&&\times 
\vec[\m_1'(\bb_1\trans\x_i) 
\otimes\{\x_{Li}-E(\X_{Li}\mid\bb_1\trans\x_i)\}]\n\\
&&
+E[P_i
\vec\{\X_{Li}\m_0'(\bb_0\trans\X_i)\trans\}]\trans
\B_0 (1-t_i)\{y_{0i}-m_0(\bb_0\trans\x_i)\}\n\\
&&\left.\left.\times 
\vec[\m_0'(\bb_0\trans\x_i) 
\otimes\{\x_{Li}-E(\X_{Li}\mid\bb_0\trans\x_i)\}] \right)\right\}.\n 
\ee 
\end{Theorem}
When $\wh D_{IMP}$ is not consistent due to misspecification of at
least one of the treatment mean models  
$m_1(\cdot)$ and $m_0(\cdot)$, $w \rightarrow 1$, thus $\sqrt{n}(\wh D
- D) \overset{d}{\rightarrow}  
\sqrt{n}(\wh D_{\rm AIPW} - D) $.

\section{Simulation Study}\label{sec:simu}

We conducted a simulation study to compare the performance of the
estimators discussed in Section 
\ref{sec:dy}. We used sample size $n=1000$ and covariate dimension
$p=6$ with 1000 replicates.  
Specifically, the covariate vector $\X =
(X_{1}, \dots, X_{6})\trans$ is generated as follows. $X_1$ and $X_2$
are generated independently from $N(1,1)$ and $N(0,1)$ distribution,
respectively. We let $X_4 = 0.015 X_1 + u_1$, where $u_1$ is uniformly
distributed in $(-0.5,0.5)$.  
Then $X_3$ and $X_5$ are generated independently from the Bernoulli
distribution with success probabilities  
$0.5+0.05 X_2$ and $0.4 + 0.2 X_4$, respectively. We let $X_6 = 0.04
X_2 + 0.15 X_3 + 0.05 X_4 + u_2$,  
where $u_2 \sim N(0,1)$.
We set $\bb_1 =
(1,-1,1,-2,-1.5,0.5)\trans$, $\bb_0=(1,1,0,0,0,0)\trans$ and
$\ba=(-0.27,0.2,-0.15,0.05,0.15,-0.1)\trans$.

\subsection{Study 1}\label{subsec:simu1}
Our first study is designed to study the estimators when the response
and propensity score models are correctly specified. 
We generated
  the response variables based on
$ Y_1 = 0.7(\bb_1 \trans \x)^2 + \sin(\bb_1 \trans \x) + \epsilon_1$
and $ Y_0 = \bb_0 \trans \x + \epsilon_0$.
Here $\epsilon_1$ and $\epsilon_0$  are normally
distributed with mean zero and variances $0.5$ and  
$0.2$ respectively.
We let further
$\eta(\ba \trans \x)=\ba \trans \x$.  Thus, the treatment
indicator $T$ is generated from the logistic model
$
\pr(T=1|\X) = {\exp(\ba \trans \x)}/\{1+\exp(\ba \trans \x)\}.
$

We implemented the six estimators described in Section \ref{sec:dy}.
In both the nonparametric estimation of
$\eta(\cdot)$ and of
the mean functions $m_1(\cdot)$ and $m_0(\cdot)$, we used local linear regression with
Epanechnikov kernel and the bandwidth was chosen
to be $c\sigma n^{-1/5}$, where $\sigma^2$ is the estimated variances 
of the corresponding indices, while $c$ is a constant ranging from 0.1
to 3.5. When extrapolation was needed, the local linear fit at the
boundary of the support was extrapolated. 
For comparison, we also computed 
$\sumi T_iY_{1i}/(\sumi T_i)
-\sumi (1-T_i)Y_{0i}/(n-\sumi T_i)$ as the naive sample average
estimator. 

\begin{figure}[!h]
\centering
\includegraphics[width=6in]{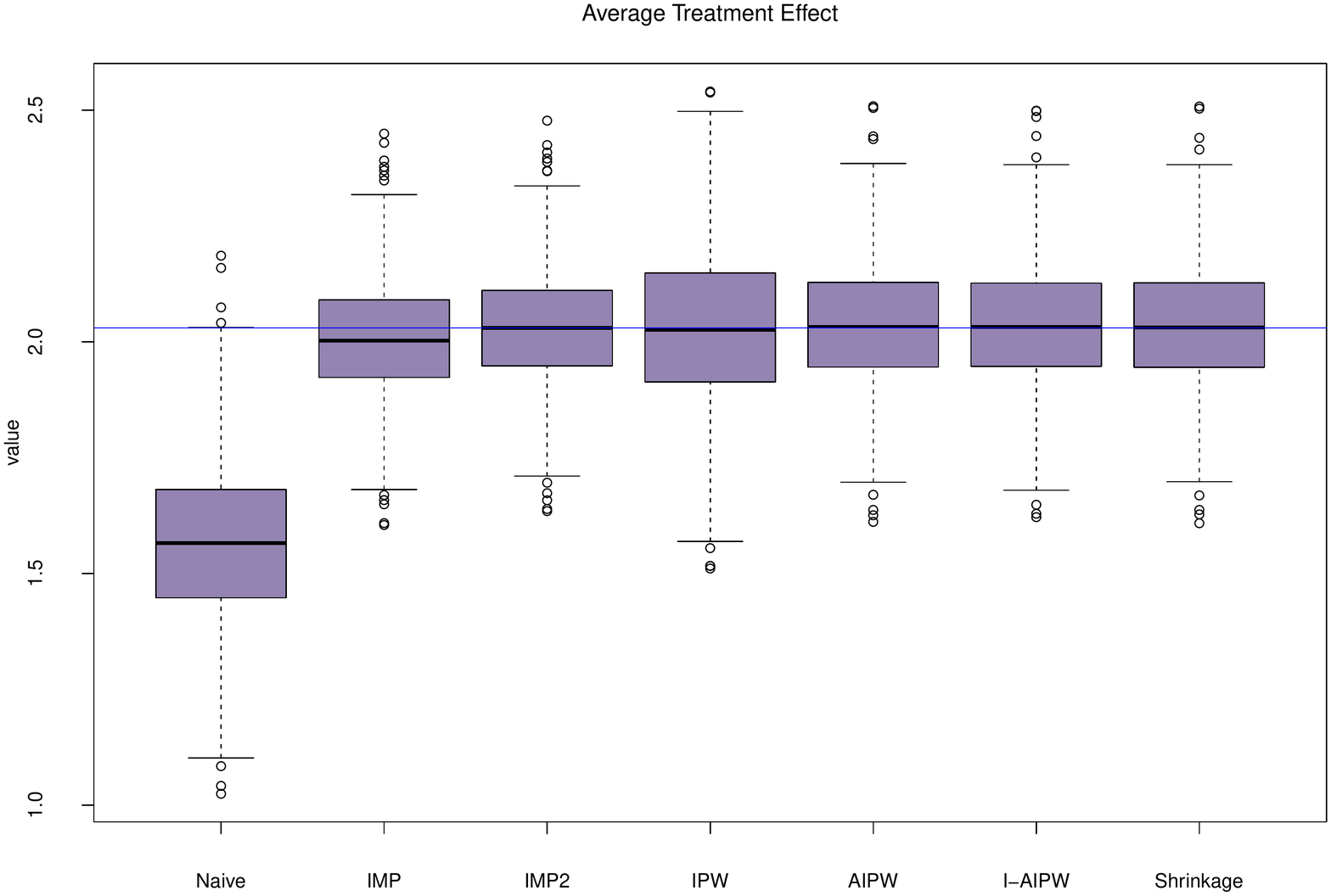}
\caption{Boxplot of Naive, IMP, IMP2, IPW, AIPW, IAIPW and
  Shrinkage estimators for Study 1.
The blue horizontal line is the true average causal
  effect, here 2.030.} 
\label{fig:avgtreat}
\end{figure}

From the results summarized in Figure \ref{fig:avgtreat} and Table
\ref{tab:avgtreat}, we can see that the naive
estimator is obviously severely biased.
As expected all six methods yield
small bias, while IMP2 and IPW provide the smallest and largest
variability and mean squared error (MSE) respectively. The estimator
shrinking IMP with AIPW improves slightly on the latter with respect
to variability and MSE. The estimated standard deviation (based on the
asymptotic developments) match fairly well the empirical variability
of the estimators.

\subsection{Study 2}\label{subsec:simu2}

The second study is designed to compare the performance of the estimators 
when the mean functions $m_1(\cdot)$ and $m_0(\cdot)$  are
misspecified. We kept the data generation procedure identical to that
of Study 1, except that we generated the response variables based
on the models
$ Y_1 = (\bb_1 \trans \x)^2 + \sin(\bb_1 \trans \x) + (\bg_1\trans\x)^2 + \epsilon_1$
and $ Y_0 = \bb_0 \trans \x + \sin(\bg_0\trans\x) +\epsilon_0$, where 
$\bg_1=(0,1,1,0,0,0)\trans$ and $\bg_0=(0,1,-0.75,0,-1,0)\trans$.
Here $\epsilon_1$ and $\epsilon_0$ are normally
distributed with mean zero and variance $0.5$ and  
$0.2$ respectively. Note that here the mean functions no longer have
the single index forms.

When we implemented the six estimators described in Section
\ref{sec:dy}, we still treated $m_1(\cdot)$ and  
$m_0(\cdot)$ as function of $\bb_1\trans\x$ and $\bb_0\trans\x$
respectively, hence the mean function models we used are misspecified.
The same nonparametric estimation procedures as in Study 1 were
used in estimating $\eta(\cdot)$, 
$m_1(\cdot)$ and $m_0(\cdot)$. 

\begin{figure}[!h]
\centering
\includegraphics[width=6in]{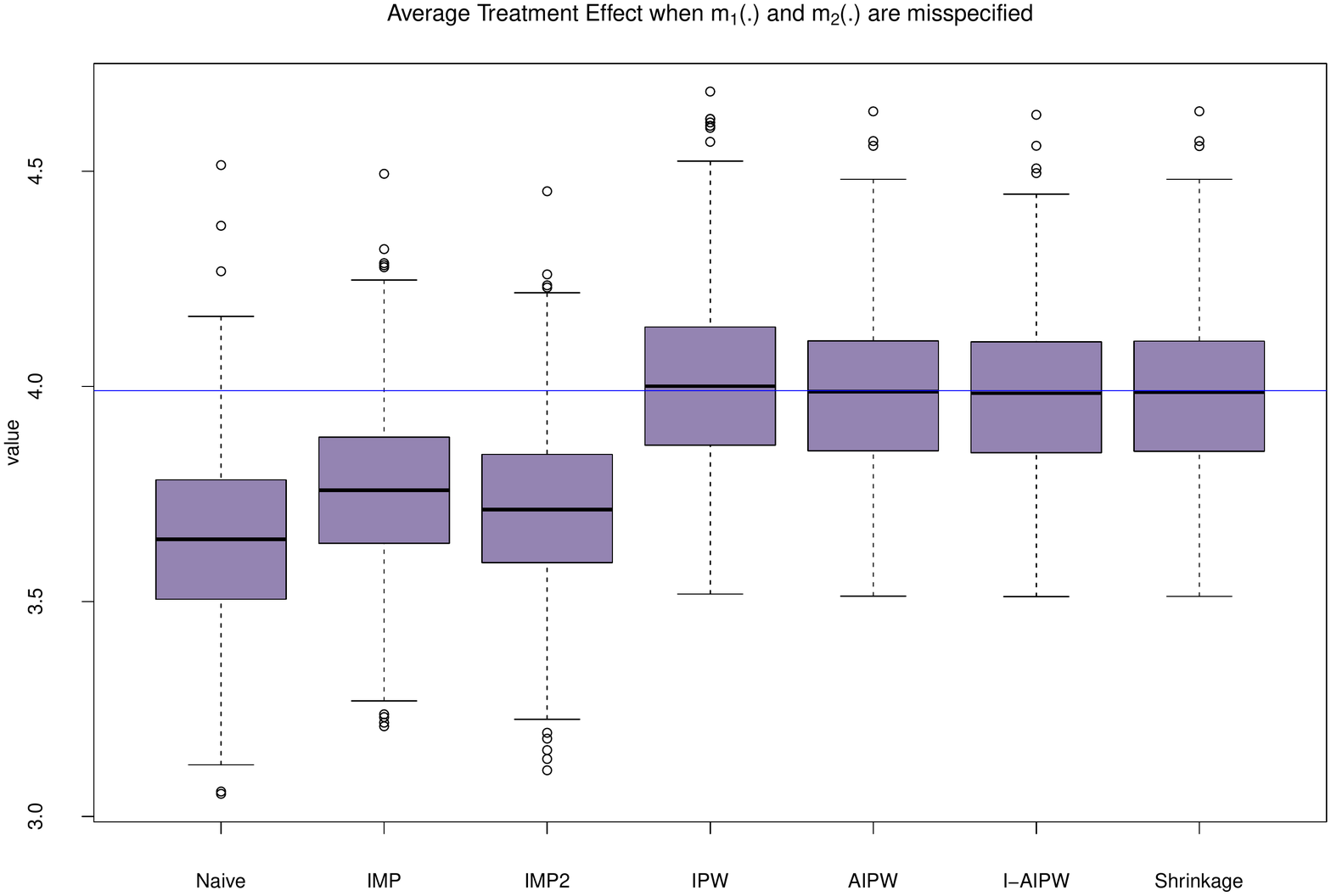}
\caption{Boxplot of Naive, IMP, IMP2, IPW, AIPW, IAIPW and
  Shrinkage estimators for Study 2, where $m_1(\cdot)$ and
  $m_0(\cdot)$ are misspecified.
The blue horizontal line is the true average causal
  effect, here 3.990.} 
\label{fig:avgtreatm}
\end{figure}

From the results in  Figure \ref{fig:avgtreatm} and Table
\ref{tab:avgtreatm}, we can see that the IMP and IMP2  
estimators are biased along with the severely biased naive estimator,
while IPW, AIPW, IAIPW and Shrinkage  
methods yield small bias, even when $m_1(\cdot)$ and
$m_0(\cdot)$ are misspecified as expected. Though IMP is biased, it provides  
the smallest variability, while IPW yields the largest
variability. Here the shrinkage estimator combining IMP and AIPW is
able to downweight IMP and inherit lower bias and variability from
AIPW. Again estimated standard deviations matches the empirical
variability of the estimators. 

\subsection{Study 3}\label{subsec:simu3}

In a third simulation study, we compare the performance of
different estimators when the model of the propensity score function
is misspecified. We followed the same data generation procedure as in
Section \ref{subsec:simu1}, but the true function inside the logistic
link  here is
$\eta(\ba \trans \x)=(\ba \trans \x)+0.45/\{(\bg\trans\x)^2+0.5\}$,
where $\bg=(1,0.5,-1,0.5,-1,-3)\trans$. So $\eta(\cdot)$ is no longer
a function of a single index.  The treatment
indicator $T$ is generated from
\bsq
\pr(T=1|\X) = \frac{\exp[(\ba \trans \x)+0.45/\{(\bg\trans\x)^2+0.5\}]}{1+
\exp[(\ba \trans \x)+0.45/ \{(\bg\trans\x)^2+0.5\}]}.
\esq

In implementing the six estimators described in Section \ref{sec:dy},
we considered $\eta(\cdot)$ as a  
function of $\ba\trans\x$ only, thus the propensity score used in
estimating the average causal effect was misspecified.  
Furthermore, we used the same nonparametric approach as in Study 1
and 2 to estimate $m_1(\cdot)$, 
$m_0(\cdot)$ and $\eta(\cdot)$.

\begin{figure}[!h]
\centering
\includegraphics[width=6in]{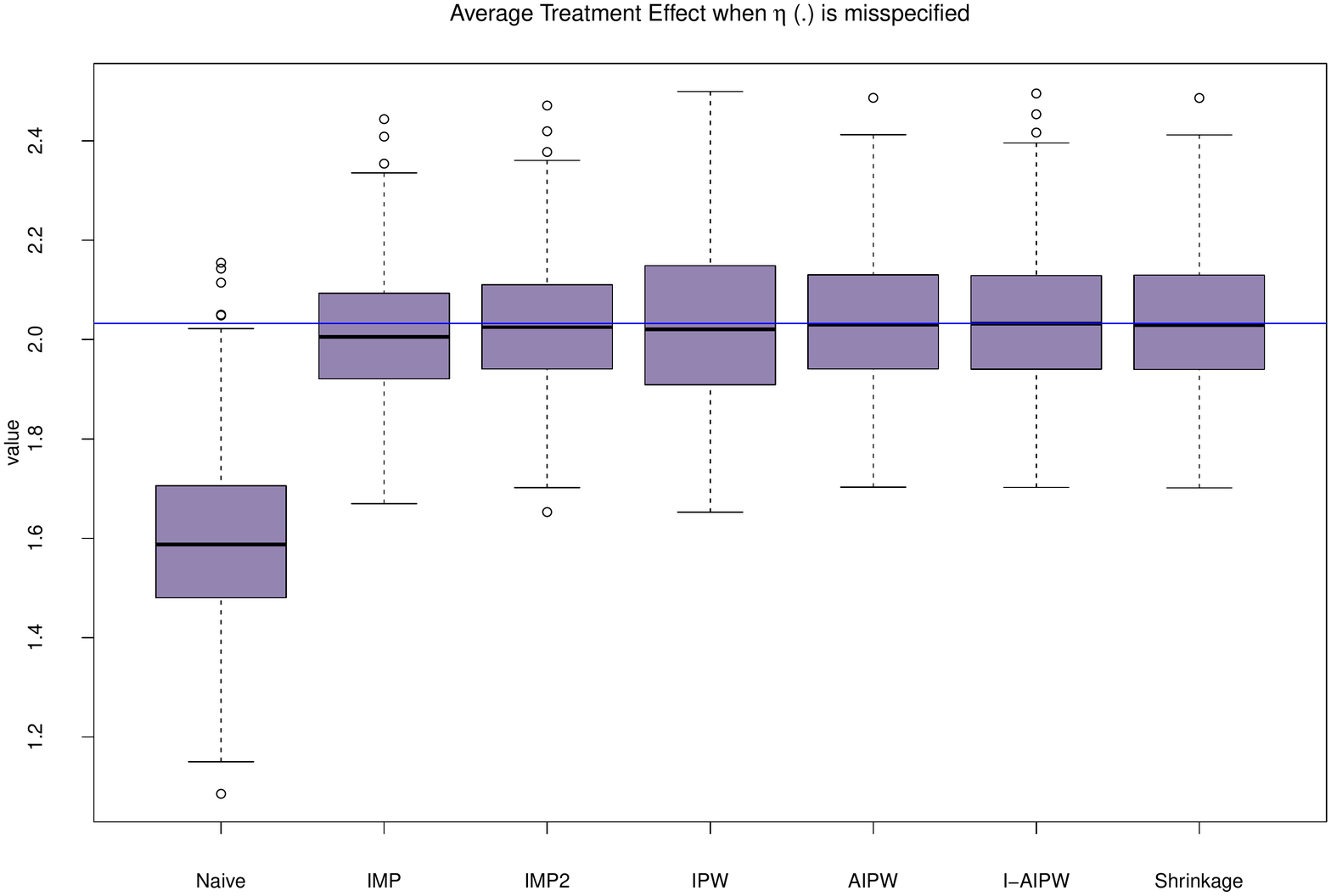}
\caption{Boxplot of Naive, IMP, IMP2, IPW, AIPW, IAIPW and
  Shrinkage estimators for Study 3, where $\eta(\cdot)$ is misspecified.
The blue horizontal line is the true average causal
  effect, here 2.033.} 
\label{fig:avgtreatp}
\end{figure}

The results in Figure \ref{fig:avgtreatp} and Table
\ref{tab:avgtreatp} show that except for the naive estimator, which 
is significantly biased,  all the six estimators yield small
biases. While the small biases of IMP, IMP2, AIPW, IAIPW and the
shrinkage estimator are within our expectation, the good performance 
of IPW is more than what the theory guarantees. 
Here IMP2 has smallest variability and MSE while IPW performs
worst. As in Study 1 both IMP and AIPW are consistent in this design
and the shrinkage estimator is again as good as AIPW. By construction,
we expect the shrinkage estimator to have lower variability in this
situation. This does not show here, probably due to the
difficulty in having precise estimates of the asymptotic variances
used to compute the shrinkage weight. On the other hand, the variance
estimates are sufficiently good to yield satisfactory empirical coverages
for the confidence intervals constructed. 

\subsection{Study 4}\label{subsec:simu4}

In this last study we consider the scenario where all models, $m_1(\cdot)$,
$m_0(\cdot)$ and $\eta(\cdot)$ are misspecified. Here the
covariate $\X$ is generated as in previous studies, the
response variables $Y_1$ and $Y_0$ are generated as in Section
\ref{subsec:simu2} and the treatment assignment as described in Section
\ref{subsec:simu3}.   
While implementing the estimators described in Section \ref{sec:dy},
we still treated $m_1(\cdot)$, $m_0(\cdot)$ and $\eta(\cdot)$ as
functions of $\bb_1\trans\x$, $\bb_0\trans\x$ and $\ba\trans\x$
respectively and  used the same nonparametric estimation procedure as in earlier sections. 

\begin{figure}[!h]
\centering
\includegraphics[width=6in]{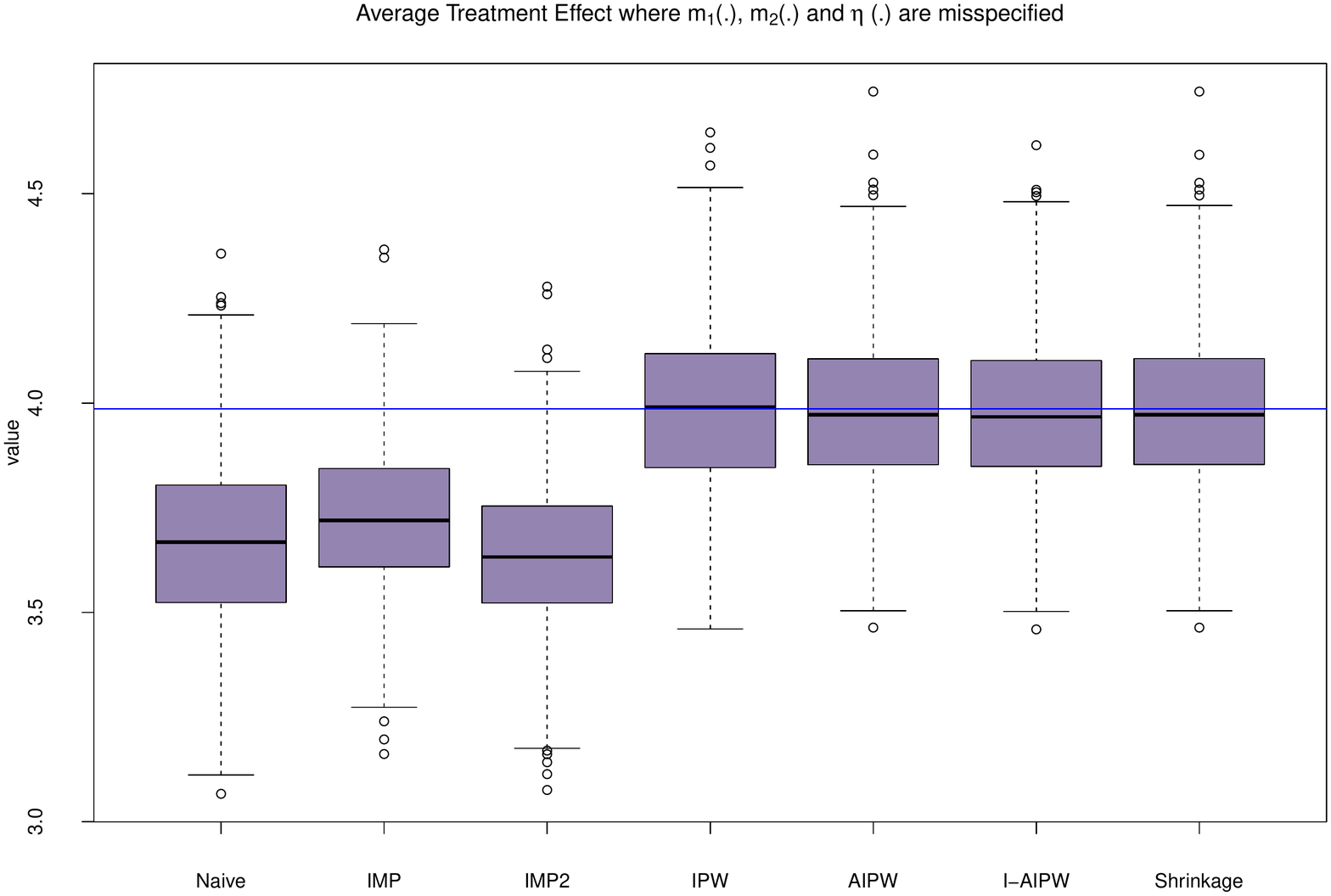}
\caption{Boxplot of Naive, IMP, IMP2, IPW, AIPW, IAIPW and
  Shrinkage estimators for Study 4,  where $m_1(\cdot)$,
  $m_0(\cdot)$
and $\eta(\cdot)$ is misspecified. The blue horizontal line is the true average causal
  effect, here 3.986.} 
\label{fig:avgtreatmp}
\end{figure}

From Figure \ref{fig:avgtreatmp} and Table \ref{tab:avgtreatmp}, we
can see that due to misspecification of the mean 
function models, IMP and IMP2 estimators are biased along with the
naive estimator. Like in Study 3, although $\eta(\cdot)$ is misspecified,   
IPW estimator yields quite small bias. Consequently, AIPW, IAIPW and
the Shrinkage estimators are also
not significantly influenced by the misspecification of response
models and the propensity score model.  IMP2 and IMP have lowest variability
followed by IAIPW and AIPW, and IPW has the
largest variance as in earlier cases. Because IMP has much larger bias than AIPW, the
shrinkage estimator mimics AIPW as the theory predicts.

\section{Data Analysis}\label{sec:data}

We now apply the methods presented to estimate the average causal effect
of maternal smoking during pregnancy on birth weight. The
data consist of birth weight (in grams) of 4642 singleton births in
Pennsylvania, USA \citep{Almond_2005}, for which several covariates
are observed: mother's age, mother's marital status, an indicator
variable for alcohol consumption during pregnancy, an indicator
variable of previous birth in which the infant died, mother's
medication, father's education, number of prenatal care visits, months
since last birth, mother's race and an indicator variable of first
born child. The data set also contains the maternal smoking habit during
pregnancy and we treat it as our treatment, $T_i$ (1=Smoking, 0=
Non-Smoking).
 This dataset was first used by \cite{Almond_2005} for
studying the economic cost of low brith weights on the
society, and was further analyzed in  \cite{CATTANEO2010} and
\cite{Liuetal2016}. The dataset can be found on
{\tt http://www.stata-press.com/data/r13/cattaneo2.dta}.  

Among the 4642 observations, 864 had smoking mothers ($T =1$) and 3778 non-smoking
($T=0$). The naive estimator (without covariate adjustment) yields an
effect of -275 grams. We used local linear regression with
Epanechnikov kernel in the nonparametric estimation of the propensity
score function, $\eta(\cdot)$ and the nonparametric estimation of the
mean functions $m_1(\cdot)$ and $m_0(\cdot)$, where the bandwidth was
selected to be $c\sigma n^{-1/5}$, $\sigma^2$ is the estimated
variance of the corresponding indices and $c$ is a constant. 
In our analysis, we find that the results are not very
  sensitive to the value of $c$, for example, when we vary $c$ from 
 from 0.1 to 95, the results hardly change. Applying the six estimators studied in Section
\ref{sec:dy} yields estimated effects of smoking within the range of
-259 to -296 gr. These are displayed in Table \ref{tab:realdata},
together with the estimated standard deviations and the $95$\% confidence
intervals. IPW stands out with an estimated effect larger than the
naive value, and this is due to some observations with propensity
scores close to zero, leading to very large weights, thereby also the
much larger standard error of IPW.  
Overall, there is evidence that smoking results in lower birth weight
given the assumption that we have observed all confounders.

\section{Discussion}\label{sec:dis}

We have introduced feasible and robust estimators of average causal
effect of a non-randomized treatment. Nuisance models are
fitted through semiparametric sufficient dimension reduction
methods. Further,  parameter estimation in these nuisance models is
locally efficient which is
important when combined with IPW and IMP estimators. AIPW
estimators are efficient and their asymptotic distribution
does not depend on the fit of the nuisance parameters as long as the
nuisance models are well specified and estimation is consistent
\citep[e.g.,][]{Farrell:15, Bellonietal}. 
The proposed shrinkage estimator
combines AIPW and IMP by improving on efficiency when the
nuisance model for the response is correctly specified. When the
latter model is misspecified the shrinkage estimator is asymptotically
equivalent to AIPW and nothing is lost eventually. Numerical
experiments show that the shrinkage estimator is at least as
performant as AIPW although no improvement could be observed over AIPW
with well specified response models, maybe due to not precise enough weights estimates
obtained with the sample size considered. 
As is the case for IMP, the shrinkage
estimator is super-efficient and its asymptotic inference is not
expected to be uniform.

\section*{Acknowledgement}
This research is supported by 
 the National Science Foundation, the National Institutes
of Health, and the Marianne and Marcus Wallenberg Foundation.

\bibliographystyle{agsm}
\bibliography{double}

\section*{Appendix}
\setcounter{subsection}{0}\renewcommand{\thesubsection}{A.\arabic{subsection}}
\subsection{Proof of IPW Properties}\label{sec:proofIPW}
\be \label{eq:a}
\wh E(Y_1)&=&n^{-1}\sumi
\frac{t_iy_i [1+\exp\{\wh\eta(\wh\ba\trans\x_i)\}]}{
  {\exp\{\wh\eta(\wh\ba\trans\x_i)\}}} \n \\
&=&n^{-1}\sumi 
t_iy_{1i} \left[
  \exp\{-\wh\eta(\ba\trans\x_i)\}+1\right]\\
&&+\left\{\frac{\partial}{\partial\vecl(\ba)\trans}\left(n^{-1}\sumi 
t_iy_{1i} \left[
  \exp\{-\eta(\ba\trans\x_i)\}+1\right]\right)+o_p(1)\right\}\vecl(\wh\ba-\ba) \n \\
&=&n^{-1}\sumi 
t_iy_{1i} \left[
  {\exp\{-\wh\eta(\ba\trans\x_i)\}}+1\right] \n \\
&&+\left\{
E\left(
T_iY_{1i} \vec\left[-
  \exp\{-\eta(\ba\trans\X_i)\}\X_{Li}\boeta'(\ba\trans\X_i)\trans
\right]\right)\right\}\trans\vecl(\wh\ba-\ba)+o_p(n^{-1/2}) \n.
\ee
Inserting (\ref{eq:useful3}),  we have that
\bse
&&\left\{
E\left(
T_iY_{1i} \vec\left[-
  \exp\{-\eta(\ba\trans\X_i)\}\X_{Li}\boeta'(\ba\trans\X_i)\trans
\right]\right)\right\}\trans\vecl(\wh\ba-\ba)\\
&=&n^{-1}\sumi\left(
E\left[
\frac{m_{1i}(\bb_1\trans\X_i)}{1+\exp\{\eta(\ba\trans\X_i)\}}
\vec\{\X_{Li}\boeta'(\ba\trans\X_i)\trans\}
\right]\right)\trans
\B \\
&&\times 
(t_i-p_i)
\vec[\{\x_{Li}-E(\X_{Li}\mid\ba\trans\x_i)\}
\boeta'(\ba\trans\x_i)\trans]+o_p(n^{-1/2}).
\ese
In addition, 
using (\ref{eq:a}) and Condition \ref{ass:bandwidth} and \ref{ass:density},
\bse
&&n^{-1}\sumi 
t_iy_{1i} \left[
  {\exp\{-\wh\eta(\ba\trans\x_i)\}}+1\right]\\
&=&n^{-1}\sumj
t_jy_{1j} \frac{n^{-1}\sumi K_{h}(\ba\trans\x_i-\ba\trans\x_j)}
       {n^{-1}\sumi t_i K_{h}(\ba\trans\x_i-\ba\trans\x_j)}\\
&=&n^{-1}\sumj
t_jy_{1j}\left[ \frac{f(\ba\trans\x_j)}
       {p_jf(\ba\trans\x_j)}
+\frac{n^{-1}\sumi K_{h}(\ba\trans\x_i-\ba\trans\x_j)
-f(\ba\trans\x_j)}{p_jf(\ba\trans\x_j)}\right.\\
&&\left.-\frac{f(\ba\trans\x_j) \{n^{-1}\sumi t_i K_{h}(\ba\trans\x_i-\ba\trans\x_j)-p_jf(\ba\trans\x_j)\}}
{\{p_jf(\ba\trans\x_j)\}^2}\right]+O_p\{h^{2m}+(nh^d)^{-1}\}\\
&=&n^{-1}\sumj
\frac{t_jy_{1j}}{p_j}\left\{ 1
+\frac{n^{-1}\sumi
  K_{h}(\ba\trans\x_i-\ba\trans\x_j)}{f(\ba\trans\x_j)}
-\frac{n^{-1}\sumi t_i
    K_{h}(\ba\trans\x_i-\ba\trans\x_j)}
{p_jf(\ba\trans\x_j)}\right\}+o_p(n^{-1/2})\\
&=&n^{-1}\sumj \frac{t_jy_{1j}}{p_j}
+n^{-2}\sumj\sumi 
\frac{t_jy_{1j} (p_j-t_i) 
    K_{h}(\ba\trans\x_i-\ba\trans\x_j)}
{p_j^2f(\ba\trans\x_j)}+o_p(n^{-1/2})\\ 
&=&n^{-1}\sumj \frac{t_jy_{1j}}{p_j}
+n^{-1}\sumj E\left\{
\frac{t_jy_{1j} (p_j-T_i) 
    K_{h}(\ba\trans\X_i-\ba\trans\x_j)}
{p_j^2f(\ba\trans\x_j)}\right\}\\
&&+n^{-1}\sumi E\left\{
\frac{T_jY_{1j} (P_j-t_i) 
    K_{h}(\ba\trans\x_i-\ba\trans\X_j)}
{P_j^2f(\ba\trans\X_j)}\right\}
-E\left\{
\frac{T_jY_{1j} (P_j-T_i) 
    K_{h}(\ba\trans\X_i-\ba\trans\X_j)}
{P_j^2f(\ba\trans\X_j)}\right\}\\
&&+o_p(n^{-1/2})\\ 
&=&n^{-1}\sumi \frac{t_iy_{1i}}{p_i}
+n^{-1}\sumi 
E\left\{ m_1(\bb_1\trans\X_i)\mid\ba\trans\x_i\right\} \left(1-\frac{t_i}
{p_i}\right)
+o_p(n^{-1/2}).
\ese
We thus obtain
\be\label{eq:ipw1}
\sqrt{n}\wh E(Y_1)
&=&n^{-1/2}\sumi \frac{t_iy_{1i}}{p_i}
+n^{-1/2}\sumi 
E\left\{ m_1(\bb_1\trans\X_i)\mid\ba\trans\x_i\right\} \left(1-\frac{t_i}
{p_i}\right)\n\\
&&+
n^{-1/2}\sumi\left(
E\left[
\frac{m_{1i}(\bb\trans\X_i)}{1+\exp\{\eta(\ba\trans\X_i)\}}
\vec\{\X_{Li}\boeta'(\ba\trans\X_i)\trans\}
\right]\right)\trans
\B \n\\
&&\times 
(t_i-p_i)
\vec[\{\x_{Li}-E(\X_{Li}\mid\ba\trans\x_i)\}
\boeta'(\ba\trans\x_i)\trans]
+o_p(1).
\ee
Similarly,
\bse 
\wh E(Y_0)&=&n^{-1}\sumi
(1-t_i)y_i [1+\exp\{\wh\eta(\wh\ba\trans\x_i)\}]\\
&=&n^{-1}\sumi 
(1-t_i)y_{0i} \left[
  \exp\{\wh\eta(\ba\trans\x_i)\}+1\right]\\
&&+\left\{\frac{\partial}{\partial\vecl(\ba)\trans}\left(n^{-1}\sumi 
(1-t_i)y_{0i} \left[
  \exp\{\eta(\ba\trans\x_i)\}+1\right]\right)+o_p(1)\right\}\vecl(\wh\ba-\ba)\\
&=&n^{-1}\sumi 
(1-t_i)y_{0i} \left[
  {\exp\{\wh\eta(\ba\trans\x_i)\}}+1\right]\\
&&+\left\{
E\left(
(1-T_i)Y_{0i} \vec\left[
  \exp\{\eta(\ba\trans\X_i)\}\X_{Li}\boeta'(\ba\trans\X_i)\trans
\right]\right)\right\}\trans\vecl(\wh\ba-\ba)+o_p(n^{-1/2}).
\ese
We further have that
\bse
&&\left\{
E\left(
(1-T_i)Y_{0i} \vec\left[
  \exp\{\eta(\ba\trans\X_i)\}\X_{Li}\boeta'(\ba\trans\X_i)\trans
\right]\right)\right\}\trans\vecl(\wh\ba-\ba)\\
&=&-n^{-1}\sumi\left(
E\left[
\frac{m_{0i}(\bb_0\trans\X_i) \exp\{\eta(\ba\trans\X_i)\}}{1+\exp\{\eta(\ba\trans\X_i)\}}
\vec\{\X_{Li}\boeta'(\ba\trans\X_i)\trans\}
\right]\right)\trans
\B \\
&&\times 
(t_i-p_i)
\vec[\{\x_{Li}-E(\X_{Li}\mid\ba\trans\x_i)\}
\boeta'(\ba\trans\x_i)\trans]+o_p(n^{-1/2}).
\ese
In addition, 
using (\ref{eq:a}) and Conditon \ref{ass:bandwidth} and \ref{ass:density},
\bse
&&n^{-1}\sumi 
(1-t_i)y_{0i} \left[
  {\exp\{\wh\eta(\ba\trans\x_i)\}}+1\right]\\
&=&n^{-1}\sumj
(1-t_j)y_{0j} \frac{n^{-1}\sumi K_{h}(\ba\trans\x_i-\ba\trans\x_j)}
       {n^{-1}\sumi (1-t_i) K_{h}(\ba\trans\x_i-\ba\trans\x_j)}\\
&=&n^{-1}\sumj
(1-t_j)y_{0j}\left[ \frac{f(\ba\trans\x_j)}
       {(1-p_j)f(\ba\trans\x_j)}
+\frac{n^{-1}\sumi K_{h}(\ba\trans\x_i-\ba\trans\x_j)
-f(\ba\trans\x_j)}{(1-p_j)f(\ba\trans\x_j)}\right.\\
&&\left.-\frac{f(\ba\trans\x_j) \{n^{-1}\sumi (1-t_i) K_{h}(\ba\trans\x_i-\ba\trans\x_j)-(1-p_j)f(\ba\trans\x_j)\}}
{\{(1-p_j)f(\ba\trans\x_j)\}^2}\right]+O_p\{h^{2m}+(nh)^{-1}\}\\
&=&n^{-1}\sumj
\frac{(1-t_j)y_{0j}}{1-p_j}\left\{ 1
+\frac{n^{-1}\sumi
  K_{h}(\ba\trans\x_i-\ba\trans\x_j)}{f(\ba\trans\x_j)}
-\frac{n^{-1}\sumi (1-t_i)
    K_{h}(\ba\trans\x_i-\ba\trans\x_j)}
{(1-p_j)f(\ba\trans\x_j)}\right\}\\
&&+o_p(n^{-1/2})\\
&=&n^{-1}\sumj \frac{(1-t_j)y_{0j}}{1-p_j}
-n^{-2}\sumj\sumi 
\frac{(1-t_j)y_{0j} (p_j-t_i) 
    K_{h}(\ba\trans\x_i-\ba\trans\x_j)}
{(1-p_j)^2f(\ba\trans\x_j)}+o_p(n^{-1/2})\\ 
&=&n^{-1}\sumj \frac{(1-t_j)y_{0j}}{1-p_j}
-n^{-1}\sumj E\left\{
\frac{(1-t_j)y_{0j} (p_j-T_i) 
    K_{h}(\ba\trans\X_i-\ba\trans\x_j)}
{(1-p_j)^2f(\ba\trans\x_j)}\right\}\\
&&-n^{-1}\sumi E\left\{
\frac{(1-T_j)Y_{0j} (P_j-t_i) 
    K_{h}(\ba\trans\x_i-\ba\trans\X_j)}
{(1-P_j)^2f(\ba\trans\X_j)}\right\}\\
&&+E\left\{
\frac{(1-T_j)Y_{0j} (P_j-T_i) 
    K_{h}(\ba\trans\X_i-\ba\trans\X_j)}
{(1-P_j)^2f(\ba\trans\X_j)}\right\}+o_p(n^{-1/2})\\ 
&=&n^{-1}\sumi \frac{(1-t_i)y_{0i}}{1-p_i}
+n^{-1}\sumi 
E\left\{ m_0(\bb_0\trans\X_i)\mid\ba\trans\x_i\right\} \left(\frac{t_i-p_i}
{1-p_i}\right)
+o_p(n^{-1/2}).
\ese
We thus obtain
\be\label{eq:ipw0}
\sqrt{n}\wh E(Y_0)
&=&n^{-1/2}\sumi \frac{(1-t_i)y_{0i}}{1-p_i}
+n^{-1/2}\sumi 
E\left\{ m_0(\bb_0\trans\X_i)\mid\ba\trans\x_i\right\} \left(\frac{t_i-p_i}
{1-p_i}\right)\n\\
&&-
n^{-1/2}\sumi\left(
E\left[
\frac{m_{0i}(\bb_0\trans\X_i) \exp\{\eta(\ba\trans\X_i)\}}{1+\exp\{\eta(\ba\trans\X_i)\}}
\vec\{\X_{Li}\boeta'(\ba\trans\X_i)\trans\}
\right]\right)\trans
\B \n\\
&&\times 
(t_i-p_i)
\vec[\{\x_{Li}-E(\X_{Li}\mid\ba\trans\x_i)\}
\boeta'(\ba\trans\x_i)\trans]
+o_p(1).
\ee
Combining the results of $\wh E(Y_1)$ and $\wh E(Y_0)$, we get
\bse
&&\sqrt{n}[\{\wh E(Y_1)-\wh E(Y_0)\}-\{E(Y_1)-E(Y_0)\}]\\
&=&n^{-1/2}\sumi \left\{\frac{t_iy_{1i}}{p_i}-E(Y_1)-\frac{(1-t_i)y_{0i}}{1-p_i}+E(Y_0)\right\}\\
&&+n^{-1/2}\sumi \left(1-\frac{t_i}
{p_i}\right)
E\left\{ m_1(\bb_1\trans\X_i)\mid\ba\trans\x_i\right\} 
-n^{-1/2}\sumi \left(\frac{t_i-p_i}
{1-p_i}\right)
E\left\{ m_0(\bb_0\trans\X_i)\mid\ba\trans\x_i\right\} \\
&&+
n^{-1/2}\sumi\left(
E\left[
\frac{m_{1i}(\bb_1\trans\X_i)+\exp\{\eta(\ba\trans\X_i)\} m_{0i}(\bb_0\trans\X_i)}{1+\exp\{\eta(\ba\trans\X_i)\}}
\vec\{\X_{Li}\boeta'(\ba\trans\X_i)\trans\}
\right]\right)\trans
\B \\
&&\times 
(t_i-p_i)
\vec[\{\x_{Li}-E(\X_{Li}\mid\ba\trans\x_i)\}
\boeta'(\ba\trans\x_i)\trans]
+o_p(1).
\ese

\qed
\subsection{Proof of Properties of AIPW}\label{sec:proofAIPW}
\bse
&&\sqrt{n}\{\wh E(Y_1)-E(Y_1)\}\\
&=& n^{-1/2}\sumi \left\{t_iy_i [1+\exp\{-\wh\eta(\wh\ba\trans\x_i)\}]-E(Y_1)
+
\left(1-t_i [1+\exp\{-\wh\eta(\wh\ba\trans\x_i)\}]\right)
\wh m_1(\wh\bb_1\trans\x_i)\right\}\\
&=&n^{-1/2}\sumi \left\{
t_iy_{1i} [1+\exp\{-\wh\eta(\ba\trans\x_i)\}]-E(Y_1)+
\left(1-t_i [1+\exp\{-\wh\eta(\ba\trans\x_i)\}]\right)
\wh m_1(\wh\bb_1\trans\x_i)\right\}\\
&&+\left[n^{-1}\sumi t_i\{y_{1i}-\wh m_1(\wh\bb_1\trans\x_i)\}\frac{\partial 
  \exp\{-\wh\eta(\ba\trans\x_i)\}}{\partial\vecl(\ba)\trans}+o_p(1)\right]
\sqrt{n}
\vecl(\wh\ba-\ba)
\\
&=&n^{-1/2}\sumi \left(\{y_{1i}-\wh m_1(\wh\bb_1\trans\x_i)\}
t_i [1+\exp\{-\wh\eta(\ba\trans\x_i)\}]+\wh m_1(\wh\bb_1\trans\x_i)-E(Y_1)\right)\\
&&-E\left[ \{Y_{1i}-m_1(\bb_1\trans\X_i)\} T_i\exp\{-\eta(\ba\trans\X_i)\}\vec\{\X_{Li}\boeta'(\ba\trans\X_i)\trans\}
\right]
\sqrt{n}
\vecl(\wh\ba-\ba)
+o_p(1)\\
&=&n^{-1/2}\sumi \left(\{y_{1i}-\wh m_1(\bb_1\trans\x_i)\}
t_i [1+\exp\{-\wh\eta(\ba\trans\x_i)\}]+\wh
m_1(\bb_1\trans\x_i)-E(Y_1)\right)\\
&&+\left[
E \left\{\frac{\partial m_1(\bb_1\trans\X_i) }{\partial\vecl(\bb_1)\trans}
(1-T_i
[1+\exp\{-\eta(\ba\trans\X_i)\}])\right\}+o_p(1)\right]\sqrt{n}\vecl(\wh\bb_1-\bb_1)\\
&&-\D_1\sqrt{n}\vecl(\wh\ba-\ba)+o_p(1)\\
&=&n^{-1/2}\sumi \left(\{y_{1i}-\wh m_1(\bb_1\trans\x_i)\}\left(
t_i
[1+\exp\{-\wh\eta(\ba\trans\x_i)\}]-1\right)+y_{1i}-E(Y_1)\right)\\
&&+\C_1\sqrt{n}\vecl(\wh\bb_1-\bb_1)-\D_1\sqrt{n}\vecl(\wh\ba-\ba)+o_p(1)\\
&=&n^{-1/2}\sumi \left\{\{y_{1i}-m_1(\bb_1\trans\x_i)\}\left(
t_i
[1+\exp\{-\eta(\ba\trans\x_i)\}]-1\right)+y_{1i}-E(Y_1)\right\}\\
&&-\C_1\B_1n^{-1/2}\sumi t_i\{y_{1i}-m_1(\bb_1\trans\x_i)\}
\vec[\m_1'(\bb_1\trans\x_i) 
\otimes\{\x_{Li}-E(\X_{Li}\mid\bb_1\trans\x_i)\}]\\
&&+\D_1
\B n^{-1/2}\sumi
(t_i-p_i)
\vec[\{\x_{Li}-E(\X_{Li}\mid\ba\trans\x_i)\}
\boeta'(\ba\trans\x_i)\trans]
+o_p(1)\\ 
&=&n^{-1/2}\sumi \left(\{y_{1i}-m_1(\bb_1\trans\x_i)\}
t_i
[1+\exp\{-\eta(\ba\trans\x_i)\}]+m_1(\bb_1\trans\x_i)-E(Y_1)\right.\\
&&-\C_1\B_1t_i\{y_{1i}-m_1(\bb_1\trans\x_i)\}
\vec[\m_1'(\bb_1\trans\x_i) 
\otimes\{\x_{Li}-E(\X_{Li}\mid\bb_1\trans\x_i)\}]\\
&&\left.+\D_1\B (t_i-p_i)
\vec[\{\x_{Li}-E(\X_{Li}\mid\ba\trans\x_i)\}
\boeta'(\ba\trans\x_i)\trans]
\right)+o_p(1).
\ese
Similarly,
\bse
&&\sqrt{n}\{\wh E(Y_0)-E(Y_0)\}\\
&=& n^{-1/2}\sumi \left\{(1-t_i)y_i [1+\exp\{\wh\eta(\wh\ba\trans\x_i)\}]-E(Y_0)
+
\left(1-(1-t_i) [1+\exp\{\wh\eta(\wh\ba\trans\x_i)\}]\right)
\wh m_0(\wh\bb_0\trans\x_i)\right\}\\
&=&n^{-1/2}\sumi \left\{
(1-t_i)y_{0i} [1+\exp\{\wh\eta(\ba\trans\x_i)\}]-E(Y_0)+
\left(1-(1-t_i) [1+\exp\{\wh\eta(\ba\trans\x_i)\}]\right)
\wh m_0(\wh\bb_0\trans\x_i)\right\}\\
&&+\left[n^{-1}\sumi(1-t_i)\{y_{0i}-\wh m_0(\wh\bb_0\trans\x_i)\}\frac{\partial 
  \exp\{\wh\eta(\ba\trans\x_i)\}}{\partial\vecl(\ba)\trans}+o_p(1)\right]
\sqrt{n}\vecl(\wh\ba-\ba)\\
&=&n^{-1/2}\sumi \left(\{y_{0i}-\wh m_0(\wh\bb_0\trans\x_i)\}
(1-t_i) [1+\exp\{\wh\eta(\ba\trans\x_i)\}]+\wh m_0(\wh\bb_0\trans\x_i)-E(Y_0)\right)\\
&&+E\left[ \{Y_{0i}-m_0(\bb_0\trans\X_i)\} (1-T_i)\exp\{\eta(\ba\trans\X_i)\}\vec\{\X_{Li}\boeta'(\ba\trans\X_i)\trans\}
\right]\sqrt{n}\vecl(\wh\ba-\ba)
+o_p(1)\\
&=&n^{-1/2}\sumi \left(\{y_{0i}-\wh m_0(\bb_0\trans\x_i)\}
(1-t_i) [1+\exp\{\wh\eta(\ba\trans\x_i)\}]+\wh
m_0(\bb_0\trans\x_i)-E(Y_0)\right)\\
&&+\left[
E \left\{\frac{\partial m_0(\bb_0\trans\X_i) }{\partial\vecl(\bb_0)\trans}
(1-(1-T_i)
[1+\exp\{\eta(\ba\trans\X_i)\}])\right\}+o_p(1)\right]\sqrt{n}\vecl(\wh\bb_1-\bb_1)\\
&&+\D_0\sqrt{n}\vecl(\wh\ba-\ba)+o_p(1)\\
&=&n^{-1/2}\sumi \left(\{y_{0i}-\wh m_0(\bb_0\trans\x_i)\}\left(
(1-t_i)
[1+\exp\{\wh\eta(\ba\trans\x_i)\}]-1\right)+y_{0i}-E(Y_0)\right)\\
&&+\C_0\sqrt{n}\vecl(\wh\bb_0-\bb_0)+\D_0\sqrt{n}\vecl(\wh\ba-\ba)+o_p(1)\\
&=&n^{-1/2}\sumi \left\{\{y_{0i}-m_0(\bb_0\trans\x_i)\}\left(
(1-t_i)
[1+\exp\{\eta(\ba\trans\x_i)\}]-1\right)+y_{0i}-E(Y_0)\right\}\\
&&-\C_0\B_0
 n^{-1/2}\sumi (1-t_i)\{y_{0i}-m_0(\bb_0\trans\x_i)\}
\vec[\m_0'(\bb_0\trans\x_i) 
\otimes\{\x_{Li}-E(\X_{Li}\mid\bb_0\trans\x_i)\}]\\
&&-\D_0
\B n^{-1/2}\sumi
(t_i-p_i)
\vec[\{\x_{Li}-E(\X_{Li}\mid\ba\trans\x_i)\}
\boeta'(\ba\trans\x_i)\trans]
+o_p(1)\\ 
&=&n^{-1/2}\sumi \left(\{y_{0i}-m_0(\bb_0\trans\x_i)\}
(1-t_i)
[1+\exp\{\eta(\ba\trans\x_i)\}]+m_0(\bb_0\trans\x_i)-E(Y_0)\right.\\
&&-\C_0\B_0
 (1-t_i)\{y_{0i}-m_0(\bb_0\trans\x_i)\}
\vec[\m_0'(\bb_0\trans\x_i) 
\otimes\{\x_{Li}-E(\X_{Li}\mid\bb_0\trans\x_i)\}]\\
&&\left.-\D_0\B 
(t_i-p_i)
\vec[\{\x_{Li}-E(\X_{Li}\mid\ba\trans\x_i)\}
\boeta'(\ba\trans\x_i)\trans]\right)
+o_p(1).
\ese

Combining the above results, we get
\bse
&&\sqrt{n}[\{\wh E(Y_1)-\wh E(Y_0)\}-\{E(Y_1)-E(Y_0)\}]\\
&=&
n^{-1/2}\sumi \left(\{y_{1i}-m_1(\bb_1\trans\x_i)\}
t_i
[1+\exp\{-\eta(\ba\trans\x_i)\}]+m_1(\bb_1\trans\x_i)-E(Y_1)\right.\n\\
&&-\C_1\B_1t_i\{y_{1i}-m_1(\bb_1\trans\x_i)\}
\vec[\m_1'(\bb_1\trans\x_i) 
\otimes\{\x_{Li}-E(\X_{Li}\mid\bb_1\trans\x_i)\}]\n\\
&&+\D_1\B (t_i-p_i)
\vec[\{\x_{Li}-E(\X_{Li}\mid\ba\trans\x_i)\}
\boeta'(\ba\trans\x_i)\trans]\n\\
&&- \{y_{0i}-m_0(\bb_0\trans\x_i)\}
(1-t_i)
[1+\exp\{\eta(\ba\trans\x_i)\}]-m_0(\bb_0\trans\x_i)+E(Y_0)\n\\
&&+\C_0\B_0
 (1-t_i)\{y_{0i}-m_0(\bb_0\trans\x_i)\}
\vec[\m_0'(\bb_0\trans\x_i) 
\otimes\{\x_{Li}-E(\X_{Li}\mid\bb_0\trans\x_i)\}]\n\\
&&\left.+\D_0\B 
(t_i-p_i)
\vec[\{\x_{Li}-E(\X_{Li}\mid\ba\trans\x_i)\}
\boeta'(\ba\trans\x_i)\trans]\right)
+o_p(1).\n
\ese
\qed
\subsection{Proof of Properties of IMP}\label{sec:proofIMP}
Using similar analysis as before, we get
\bse 
&&\sqrt{n}\{\wh E(Y_1)-E(Y_1)\}\\
&=&n^{-1/2}\sumi \left\{t_iy_i+(1-t_i) 
\wh m_1(\wh\bb_1\trans\x_i)-E(Y_1) \right\}\\
&=&n^{-1/2}\sumi \left\{t_iy_{1i}+(1-t_i) 
\wh m_1(\bb_1\trans\x_i)-E(Y_1) \right\}\\
&&+\left\{n^{-1}\sumi (1-t_i) 
\frac{\partial\wh
  m_1(\bb_1\trans\x_i)}{\partial\vecl(\bb_1)\trans}+o_p(1)\right\}\sqrt{n}\vecl(\wh\bb_1-\bb_1)\\
&=&n^{-1/2}\sumi \left\{t_iy_{1i}+(1-t_i) 
\wh m_1(\bb_1\trans\x_i)-E(Y_1) \right\}\\
&&+E[(1-P_i) \vec\{\X_{Li}\m_1'(\bb_1\trans\X_i)\trans\}]\trans\sqrt{n}\vecl(\wh\bb_1-\bb_1)+o_p(1).
\ese
We further have that
\bse
&&E[(1-P_i)
\vec\{\X_{Li}\m_1'(\bb_1\trans\X_i)\trans\}]\trans\sqrt{n}\vecl(\wh\bb_1-\bb_1)\\
&=&-n^{-1/2}\sumi E[(1-P_i)
\vec\{\X_{Li}\m_1'(\bb_1\trans\X_i)\trans\}]\trans
\B_1\\
&&\times t_i\{y_{1i}-m_1(\bb_1\trans\x_i)\}
\vec[\m_1'(\bb_1\trans\x_i) 
\otimes\{\x_{Li}-E(\X_{Li}\mid\bb_1\trans\x_i)\}]+o_p(1).
\ese
On the other hand, 
\bse
&&n^{-1/2}\sumi \left\{t_iy_{1i}+(1-t_i) 
\wh m_1(\bb_1\trans\x_i)-E(Y_1) \right\}\\
&=&n^{-1/2}\sumi \left\{t_iy_{1i}+(1-t_i) 
\frac{n^{-1}\sumj t_jy_{1j}K_h(\bb_1\trans\x_j-\bb_1\trans\x_i)}
{n^{-1}\sumj t_jK_h(\bb_1\trans\x_j-\bb_1\trans\x_i)}-E(Y_1) \right\}\\
&=&n^{-1/2}\sumi \left[t_iy_{1i}+(1-t_i) \left\{
\frac{E(T_iY_{1i}\mid\bb_1\trans\x_i)f(\bb_1\trans\x_i)}
{E( T_i\mid\bb_1\trans\x_i)f(\bb_1\trans\x_i)}\right.\right.\\
&&
+\frac{n^{-1}\sumj t_jy_{1j}K_h(\bb_1\trans\x_j-\bb_1\trans\x_i)- E(T_iY_{1i}\mid\bb_1\trans\x_i)f(\bb_1\trans\x_i)}
{E( T_i\mid\bb_1\trans\x_i)f(\bb_1\trans\x_i)}\\
&&\left.\left.-E(T_iY_{1i}\mid\bb_1\trans\x_i)f(\bb_1\trans\x_i)\frac{n^{-1}\sumj t_jK_h(\bb_1\trans\x_j-\bb_1\trans\x_i)- E(T_i\mid\bb_1\trans\x_i)f(\bb_1\trans\x_i)}
{\{E( T_i\mid\bb_1\trans\x_i)f(\bb_1\trans\x_i)\}^2}
\right\}-E(Y_1) \right]\\
&&+O_p(n^{1/2}h^{2m}+n^{-1/2}h^{-d})\\
&=&n^{-1/2}\sumi \left[t_iy_{1i}+(1-t_i) \left\{
m_1(\bb_1\trans\x_i) 
+\frac{n^{-1}\sumj t_j\{y_{1j}-m_1(\bb_1\trans\x_i)\}
K_h(\bb_1\trans\x_j-\bb_1\trans\x_i)}
{p_if(\bb_1\trans\x_i)}
\right\}-E(Y_1) \right]\\
&&+o_p(1)\\
&=&n^{-1/2}\sumi \left\{t_iy_{1i}+
(1-t_i)  m_1(\bb_1\trans\x_i) -E(Y_1)\right\}\\
&&+n^{-3/2}\sumi \sumj
\frac{(1-t_i)  t_j\{y_{1j}-m_1(\bb_1\trans\x_i)\}
K_h(\bb_1\trans\x_j-\bb_1\trans\x_i)}
{p_if(\bb_1\trans\x_i)} +o_p(1)\\
&=&n^{-1/2}\sumi \left\{t_iy_{1i}+
(1-t_i)  m_1(\bb_1\trans\x_i) -E(Y_1)\right\}\\
&&+n^{-1/2}\sumi E\left\{
\frac{(1-t_i)  T_j\{Y_{1j}-m_1(\bb_1\trans\x_i)\}
K_h(\bb_1\trans\X_j-\bb_1\trans\x_i)}
{p_if(\bb_1\trans\x_i)}\right\}\\
&&+n^{-1/2}\sumj E\left\{
\frac{(1-T_i)  t_j\{y_{1j}-m_1(\bb_1\trans\X_i)\}
K_h(\bb_1\trans\x_j-\bb_1\trans\X_i)}
{P_if(\bb_1\trans\X_i)} \right\}\\
&&-n^{1/2} E\left\{
\frac{(1-T_i)  T_j\{Y_{1j}-m_1(\bb_1\trans\X_i)\}
K_h(\bb_1\trans\X_j-\bb_1\trans\X_i)}
{P_if(\bb_1\trans\X_i)} \right\}+o_p(1)\\
&=&n^{-1/2}\sumi \left\{t_iy_{1i}+
(1-t_i)  m_1(\bb_1\trans\x_i) -E(Y_1)\right\}\\
&&+n^{-1/2}\sumj 
E[\exp\{-\eta(\ba\trans\X_j)\}\mid\bb_1\trans\x_j]t_j\{y_{1j}-m_1(\bb_1\trans\x_j)\}
+o_p(1).
\ese
Combining the above results, we get
\bse
&&\sqrt{n}\{\wh E(Y_1)-E(Y_1)\}\\
&=&n^{-1/2}\sumi \left\{t_iy_{1i}+
(1-t_i)  m_1(\bb_1\trans\x_i) -E(Y_1)\right\}\\
&&+n^{-1/2}\sumi 
E[\exp\{-\eta(\ba\trans\X_i)\}\mid\bb_1\trans\x_i]t_i\{y_{1i}-m_1(\bb_1\trans\x_i)\}\\
&&-n^{-1/2}\sumi E[(1-P_i)
\vec\{\X_{Li}\m_1'(\bb_1\trans\X_i)\trans\}]\trans
\B_1\\
&&\times t_i\{y_{1i}-m_1(\bb_1\trans\x_i)\}
\vec[\m_1'(\bb_1\trans\x_i) 
\otimes\{\x_{Li}-E(\X_{Li}\mid\bb_1\trans\x_i)\}]+o_p(1).
\ese
Similarly,
\bse 
&&\sqrt{n}\{\wh E(Y_0)-E(Y_0)\}\\
&=&n^{-1/2}\sumi \left\{(1-t_i)y_i+t_i
\wh m_0(\wh\bb_0\trans\x_i)-E(Y_0) \right\}\\
&=&n^{-1/2}\sumi \left\{(1-t_i)y_{0i}+t_i
\wh m_0(\bb_0\trans\x_i)-E(Y_0) \right\}\\
&&+\left\{n^{-1}\sumi t_i
\frac{\partial\wh
  m_0(\bb_0\trans\x_i)}{\partial\vecl(\bb_0)\trans}+o_p(1)\right\}\sqrt{n}\vecl(\wh\bb_0-\bb_0)\\
&=&n^{-1/2}\sumi \left\{(1-t_i)y_{0i}+t_i 
\wh m_0(\bb_0\trans\x_i)-E(Y_0) \right\}\\
&&+E[P_i \vec\{\X_{Li}\m_0'(\bb_0\trans\X_i)\trans\}]\trans\sqrt{n}\vecl(\wh\bb_0-\bb_0)+o_p(1).
\ese
We have that
\bse
&&E[P_i
\vec\{\X_{Li}\m_0'(\bb_0\trans\X_i)\trans\}]\trans\sqrt{n}\vecl(\wh\bb_0-\bb_0)\\
&=&-n^{-1/2}\sumi E[P_i
\vec\{\X_{Li}\m_0'(\bb_0\trans\X_i)\trans\}]\trans
\B_0\\
&&\times (1-t_i)\{y_{0i}-m_0(\bb_0\trans\x_i)\}
\vec[\m_0'(\bb_0\trans\x_i) 
\otimes\{\x_{Li}-E(\X_{Li}\mid\bb_0\trans\x_i)\}]+o_p(1).
\ese
On the other hand, 
\bse
&&n^{-1/2}\sumi \left\{(1-t_i)y_{0i}+t_i 
\wh m_0(\bb_0\trans\x_i)-E(Y_0) \right\}\\
&=&n^{-1/2}\sumi \left\{(1-t_i)y_{0i}+t_i
\frac{n^{-1}\sumj (1-t_j)y_{0j}K_h(\bb_0\trans\x_j-\bb_0\trans\x_i)}
{n^{-1}\sumj (1-t_j)K_h(\bb_0\trans\x_j-\bb_0\trans\x_i)}-E(Y_0) \right\}\\
&=&n^{-1/2}\sumi \left[(1-t_i)y_{0i}+t_i\left\{
\frac{E\{(1-T_i)Y_{0i}\mid\bb_0\trans\x_i\}f(\bb_0\trans\x_i)}
{E\{(1-T_i)\mid\bb_0\trans\x_i\}f(\bb_0\trans\x_i)}\right.\right.\\
&&
+\frac{n^{-1}\sumj (1-t_j)y_{0j}K_h(\bb_0\trans\x_j-\bb_0\trans\x_i)- E\{(1-T_i)Y_{0i}\mid\bb_0\trans\x_i\}f(\bb_0\trans\x_i)}
{E(1-T_i\mid\bb_0\trans\x_i)f(\bb_0\trans\x_i)}\\
&&\left.\left.-E\{(1-T_i)Y_{0i}\mid\bb_0\trans\x_i\}f(\bb_0\trans\x_i)\frac{n^{-1}\sumj (1-t_j)K_h(\bb_0\trans\x_j-\bb_0\trans\x_i)- E(1-T_i\mid\bb_0\trans\x_i)f(\bb_0\trans\x_i)}
{\{E(1- T_i\mid\bb_0\trans\x_i)f(\bb_0\trans\x_i)\}^2}
\right\}\right.\\
&&\left.-E(Y_0) \right]+O_p(n^{1/2}h^{2m}+n^{-1/2}h^{-d})\\
&=&n^{-1/2}\sumi \left[(1-t_i)y_{0i}+t_i \left\{
m_0(\bb_0\trans\x_i) 
+\frac{n^{-1}\sumj (1-t_j)\{y_{0j}-m_0(\bb_0\trans\x_i)\}
K_h(\bb_0\trans\x_j-\bb_0\trans\x_i)}
{(1-p_i)f(\bb_0\trans\x_i)}
\right\}\right.\\
&&\left.-E(Y_0) \right]+o_p(1)\\
&=&n^{-1/2}\sumi \left\{(1-t_i)y_{0i}+t_i m_0(\bb_0\trans\x_i) -E(Y_0)\right\}\\
&&+n^{-3/2}\sumi \sumj
\frac{t_i(1- t_j)\{y_{0j}-m_0(\bb_0\trans\x_i)\}
K_h(\bb_0\trans\x_j-\bb_0\trans\x_i)}
{(1-p_i)f(\bb_0\trans\x_i)} +o_p(1)\\
&=&n^{-1/2}\sumi \left\{(1-t_i)y_{0i}+
t_i  m_0(\bb_0\trans\x_i) -E(Y_0)\right\}\\
&&+n^{-1/2}\sumi E\left\{
\frac{t_i (1- T_j)\{Y_{0j}-m_0(\bb_0\trans\x_i)\}
K_h(\bb_0\trans\X_j-\bb_0\trans\x_i)}
{(1-p_i)f(\bb_0\trans\x_i)}\right\}\\
&&+n^{-1/2}\sumj E\left\{
\frac{T_i (1- t_j)\{y_{0j}-m_0(\bb_0\trans\X_i)\}
K_h(\bb_0\trans\x_j-\bb_0\trans\X_i)}
{(1-P_i)f(\bb_0\trans\X_i)} \right\}\\
&&-n^{1/2} E\left\{
\frac{T_i(1-T_j)\{Y_{0j}-m_0(\bb_0\trans\X_i)\}
K_h(\bb_0\trans\X_j-\bb_0\trans\X_i)}
{(1-P_i)f(\bb_0\trans\X_i)} \right\}+o_p(1)\\
&=&n^{-1/2}\sumi \left\{(1-t_i)y_{0i}+
t_i  m_0(\bb_0\trans\x_i) -E(Y_0)\right\}\\
&&+n^{-1/2}\sumj 
E[\exp\{\eta(\ba\trans\X_j)\}\mid\bb_0\trans\x_j](1-t_j)\{y_{0j}-m_0(\bb_0\trans\x_j)\}
+o_p(1).
\ese
Combining the above results, we get
\bse
&&\sqrt{n}\{\wh E(Y_0)-E(Y_0)\}\\
&=&n^{-1/2}\sumi \left\{(1-t_i)y_{0i}+
t_i  m_0(\bb_0\trans\x_i) -E(Y_0)\right\}\\
&&+n^{-1/2}\sumi
E[\exp\{\eta(\ba\trans\X_i)\}\mid\bb_0\trans\x_i](1-t_i)\{y_{0i}-m_0(\bb_0\trans\x_i)\}\\
&&
-n^{-1/2}\sumi E[P_i
\vec\{\X_{Li}\m_0'(\bb_0\trans\X_i)\trans\}]\trans
\B_0\\
&&\times (1-t_i)\{y_{0i}-m_0(\bb_0\trans\x_i)\}
\vec[\m_0'(\bb_0\trans\x_i) 
\otimes\{\x_{Li}-E(\X_{Li}\mid\bb_0\trans\x_i)\}] +o_p(1).
\ese
Now combining the results regarding $\wh E(Y_1)$ and $\wh E(Y_0)$, we
get
\bse
&&\sqrt{n}[\{\wh E(Y_1)-E(Y_1)\}-\{\wh E(Y_0)-E(Y_0)\}]\\
&=&n^{-1/2}\sumi \left\{t_iy_{1i}-(1-t_i)y_{0i}
+(1-t_i)  m_1(\bb_1\trans\x_i) 
-t_i  m_0(\bb_0\trans\x_i) -E(Y_1) +E(Y_0)\right\}\\
&&+n^{-1/2}\sumi 
E[\exp\{-\eta(\ba\trans\X_i)\}\mid\bb_1\trans\x_i]t_i\{y_{1i}-m_1(\bb_1\trans\x_i)\}\\
&&-n^{-1/2}\sumi
E[\exp\{\eta(\ba\trans\X_i)\}\mid\bb_0\trans\x_i](1-t_i)\{y_{0i}-m_0(\bb_0\trans\x_i)\}\\
&&-n^{-1/2}\sumi E[(1-P_i)
\vec\{\X_{Li}\m_1'(\bb_1\trans\X_i)\trans\}]\trans
\B_1\\
&&\times t_i\{y_{1i}-m_1(\bb_1\trans\x_i)\}
\vec[\m_1'(\bb_1\trans\x_i) 
\otimes\{\x_{Li}-E(\X_{Li}\mid\bb_1\trans\x_i)\}]\\
&&
+n^{-1/2}\sumi E[P_i
\vec\{\X_{Li}\m_0'(\bb_0\trans\X_i)\trans\}]\trans
\B_0\\
&&\times (1-t_i)\{y_{0i}-m_0(\bb_0\trans\x_i)\}
\vec[\m_0'(\bb_0\trans\x_i) 
\otimes\{\x_{Li}-E(\X_{Li}\mid\bb_0\trans\x_i)\}] +o_p(1)\\
&=&
n^{-1/2}\sumi \left\{
m_1(\bb_1\trans\x_i) 
- m_0(\bb_0\trans\x_i) -E(Y_1) +E(Y_0)\right\}\\
&&+n^{-1/2}\sumi 
E[1+\exp\{-\eta(\ba\trans\X_i)\}\mid\bb_1\trans\x_i]t_i\{y_{1i}-m_1(\bb_1\trans\x_i)\}\\
&&-n^{-1/2}\sumi
E[1+\exp\{\eta(\ba\trans\X_i)\}\mid\bb_0\trans\x_i](1-t_i)\{y_{0i}-m_0(\bb_0\trans\x_i)\}\\
&&-n^{-1/2}\sumi E[(1-P_i)
\vec\{\X_{Li}\m_1'(\bb_1\trans\X_i)\trans\}]\trans
\B_1\\
&&\times t_i\{y_{1i}-m_1(\bb_1\trans\x_i)\}
\vec[\m_1'(\bb_1\trans\x_i) 
\otimes\{\x_{Li}-E(\X_{Li}\mid\bb_1\trans\x_i)\}]\\
&&
+n^{-1/2}\sumi E[P_i
\vec\{\X_{Li}\m_0'(\bb_0\trans\X_i)\trans\}]\trans
\B_0\\
&&\times (1-t_i)\{y_{0i}-m_0(\bb_0\trans\x_i)\}
\vec[\m_0'(\bb_0\trans\x_i) 
\otimes\{\x_{Li}-E(\X_{Li}\mid\bb_0\trans\x_i)\}] +o_p(1).
\ese
\qed

\subsection{Proof of Properties of IMP2}\label{sec:proofIMP2}

\bse
&&n^{1/2}\{\wh E(Y_1)-E(Y_1)\}\\
&=&n^{-1/2}\sumi \left\{
\wh m_1(\wh\bb_1\trans\x_i)-E(Y_1) \right\}\\
&=&n^{-1/2}\sumi \left\{
\wh m_1(\bb_1\trans\x_i)-E(Y_1) \right\}
+\left\{n^{-1}\sumi 
\frac{\partial\wh
  m_1(\bb_1\trans\x_i)}{\partial\vecl(\bb_1)\trans}+o_p(1)\right\}\sqrt{n}\vecl(\wh\bb_1-\bb_1)\\
&=&n^{-1/2}\sumi \left\{
\wh m_1(\bb_1\trans\x_i)-E(Y_1) \right\}
+E[\vec\{\X_{Li}\m_1'(\bb_1\trans\X_i)\trans\}]\trans\sqrt{n}\vecl(\wh\bb_1-\bb_1)+o_p(1).
\ese
We further have that
\bse
&&E[
\vec\{\X_{Li}\m_1'(\bb_1\trans\X_i)\trans\}]\trans\sqrt{n}\vecl(\wh\bb_1-\bb_1)\\
&=&-n^{-1/2}\sumi E[
\vec\{\X_{Li}\m_1'(\bb_1\trans\X_i)\trans\}]\trans
\B_1\\
&&\times t_i\{y_{1i}-m_1(\bb_1\trans\x_i)\}
\vec[\m_1'(\bb_1\trans\x_i) 
\otimes\{\x_{Li}-E(\X_{Li}\mid\bb_1\trans\x_i)\}]+o_p(1).
\ese
On the other hand, 
\bse
&&n^{-1/2}\sumi \left\{
\wh m_1(\bb_1\trans\x_i)-E(Y_1) \right\}\\
&=&n^{-1/2}\sumi \left\{
\frac{n^{-1}\sumj t_jy_{1j}K_h(\bb_1\trans\x_j-\bb_1\trans\x_i)}
{n^{-1}\sumj t_jK_h(\bb_1\trans\x_j-\bb_1\trans\x_i)}-E(Y_1) \right\}\\
&=&n^{-1/2}\sumi \left\{
\frac{E(T_iY_{1i}\mid\bb_1\trans\x_i)f(\bb_1\trans\x_i)}
{E( T_i\mid\bb_1\trans\x_i)f(\bb_1\trans\x_i)}
+\frac{n^{-1}\sumj t_jy_{1j}K_h(\bb_1\trans\x_j-\bb_1\trans\x_i)- E(T_iY_{1i}\mid\bb_1\trans\x_i)f(\bb_1\trans\x_i)}
{E( T_i\mid\bb_1\trans\x_i)f(\bb_1\trans\x_i)}\right.\\
&&\left.-E(T_iY_{1i}\mid\bb_1\trans\x_i)f(\bb_1\trans\x_i)\frac{n^{-1}\sumj t_jK_h(\bb_1\trans\x_j-\bb_1\trans\x_i)- E(T_i\mid\bb_1\trans\x_i)f(\bb_1\trans\x_i)}
{\{E( T_i\mid\bb_1\trans\x_i)f(\bb_1\trans\x_i)\}^2}
-E(Y_1) \right\}\\
&&+O_p(n^{1/2}h^{2m}+n^{-1/2}h^{-d})\\
&=&n^{-1/2}\sumi  \left\{
m_1(\bb_1\trans\x_i) 
+\frac{n^{-1}\sumj t_j\{y_{1j}-m_1(\bb_1\trans\x_i)\}
K_h(\bb_1\trans\x_j-\bb_1\trans\x_i)}
{p_if(\bb_1\trans\x_i)}-E(Y_1) \right\}+o_p(1)\\
&=&n^{-1/2}\sumi \left\{m_1(\bb_1\trans\x_i) -E(Y_1)\right\}
+n^{-3/2}\sumi \sumj
\frac{t_j\{y_{1j}-m_1(\bb_1\trans\x_i)\}
K_h(\bb_1\trans\x_j-\bb_1\trans\x_i)}
{p_if(\bb_1\trans\x_i)} +o_p(1)\\
&=&n^{-1/2}\sumi \left\{m_1(\bb_1\trans\x_i) -E(Y_1)\right\}\\
&&+n^{-1/2}\sumi E\left\{
\frac{ T_j\{Y_{1j}-m_1(\bb_1\trans\x_i)\}
K_h(\bb_1\trans\X_j-\bb_1\trans\x_i)}
{p_if(\bb_1\trans\x_i)}\right\}\\
&&+n^{-1/2}\sumj E\left\{
\frac{t_j\{y_{1j}-m_1(\bb_1\trans\X_i)\}
K_h(\bb_1\trans\x_j-\bb_1\trans\X_i)}
{P_if(\bb_1\trans\X_i)} \right\}\\
&&-n^{1/2} E\left\{
\frac{T_j\{Y_{1j}-m_1(\bb_1\trans\X_i)\}
K_h(\bb_1\trans\X_j-\bb_1\trans\X_i)}
{P_if(\bb_1\trans\X_i)} \right\}+o_p(1)\\
&=&n^{-1/2}\sumi \left\{m_1(\bb_1\trans\x_i) -E(Y_1)\right\}\\
&&+n^{-1/2}\sumj 
E[1+\exp\{-\eta(\ba\trans\X_j)\}\mid\bb_1\trans\x_j]t_j\{y_{1j}-m_1(\bb_1\trans\x_j)\}
+o_p(1).
\ese
Combining the above results, we get
\bse
&&\sqrt{n}\{\wh E(Y_1)-E(Y_1)\}\\
&=&n^{-1/2}\sumi \left\{m_1(\bb_1\trans\x_i) -E(Y_1)\right\}
+n^{-1/2}\sumi 
E[1+\exp\{-\eta(\ba\trans\X_i)\}\mid\bb_1\trans\x_i]t_i\{y_{1i}-m_1(\bb_1\trans\x_i)\}\\
&&-n^{-1/2}\sumi E[
\vec\{\X_{Li}\m_1'(\bb_1\trans\X_i)\trans\}]\trans
\B_1\\
&&\times t_i\{y_{1i}-m_1(\bb_1\trans\x_i)\}
\vec[\m_1'(\bb_1\trans\x_i) 
\otimes\{\x_{Li}-E(\X_{Li}\mid\bb_1\trans\x_i)\}]+o_p(1).
\ese
Similarly,
\bse 
&&\sqrt{n}\{\wh E(Y_0)-E(Y_0)\}\\
&=&n^{-1/2}\sumi \left\{
\wh m_0(\wh\bb_0\trans\x_i)-E(Y_0) \right\}\\
&=&n^{-1/2}\sumi \left\{
\wh m_0(\bb_0\trans\x_i)-E(Y_0) \right\}
+\left\{n^{-1}\sumi 
\frac{\partial\wh
  m_0(\bb_0\trans\x_i)}{\partial\vecl(\bb_0)\trans}+o_p(1)\right\}\sqrt{n}\vecl(\wh\bb_0-\bb_0)\\
&=&n^{-1/2}\sumi \left\{
\wh m_0(\bb_0\trans\x_i)-E(Y_0) \right\}
+E[\vec\{\X_{Li}\m_0'(\bb_0\trans\X_i)\trans\}]\trans\sqrt{n}\vecl(\wh\bb_0-\bb_0)+o_p(1).
\ese
We further have that
\bse
&&E[
\vec\{\X_{Li}\m_0'(\bb_0\trans\X_i)\trans\}]\trans\sqrt{n}\vecl(\wh\bb_0-\bb_0)\\
&=&-n^{-1/2}\sumi E[
\vec\{\X_{Li}\m_0'(\bb_0\trans\X_i)\trans\}]\trans
\B_0\\
&&\times (1-t_i)\{y_{0i}-m_0(\bb_0\trans\x_i)\}
\vec[\m_0'(\bb_0\trans\x_i) 
\otimes\{\x_{Li}-E(\X_{Li}\mid\bb_0\trans\x_i)\}]+o_p(1).
\ese
On the other hand, 
\bse
&&n^{-1/2}\sumi \left\{
\wh m_0(\bb_0\trans\x_i)-E(Y_0) \right\}\\
&=&n^{-1/2}\sumi \left\{
\frac{n^{-1}\sumj (1-t_j)y_{0j}K_h(\bb_0\trans\x_j-\bb_0\trans\x_i)}
{n^{-1}\sumj (1-t_j)K_h(\bb_0\trans\x_j-\bb_0\trans\x_i)}-E(Y_0) \right\}\\
&=&n^{-1/2}\sumi \left\{
\frac{E\{(1-T_i)Y_{0i}\mid\bb_0\trans\x_i\}f(\bb_0\trans\x_i)}
{E\{(1-T_i)\mid\bb_0\trans\x_i\}f(\bb_0\trans\x_i)}\right.\\
&&
+\frac{n^{-1}\sumj (1-t_j)y_{0j}K_h(\bb_0\trans\x_j-\bb_0\trans\x_i)- E\{(1-T_i)Y_{0i}\mid\bb_0\trans\x_i\}f(\bb_0\trans\x_i)}
{E(1-T_i\mid\bb_0\trans\x_i)f(\bb_0\trans\x_i)}\\
&&\left.-E\{(1-T_i)Y_{0i}\mid\bb_0\trans\x_i\}f(\bb_0\trans\x_i)\frac{n^{-1}\sumj (1-t_j)K_h(\bb_0\trans\x_j-\bb_0\trans\x_i)- E(1-T_i\mid\bb_0\trans\x_i)f(\bb_0\trans\x_i)}
{\{E(1- T_i\mid\bb_0\trans\x_i)f(\bb_0\trans\x_i)\}^2}
\right.\\
&&\left.-E(Y_0) \right\}+O_p(n^{1/2}h^{2m}+n^{-1/2}h^{-d})\\
&=&n^{-1/2}\sumi\left\{
m_0(\bb_0\trans\x_i) 
+\frac{n^{-1}\sumj (1-t_j)\{y_{0j}-m_0(\bb_0\trans\x_i)\}
K_h(\bb_0\trans\x_j-\bb_0\trans\x_i)}
{(1-p_i)f(\bb_0\trans\x_i)}-E(Y_0) \right\}+o_p(1)\\
&=&n^{-1/2}\sumi \left\{m_0(\bb_0\trans\x_i) -E(Y_0)\right\}\\
&&+n^{-3/2}\sumi \sumj
\frac{(1- t_j)\{y_{0j}-m_0(\bb_0\trans\x_i)\}
K_h(\bb_0\trans\x_j-\bb_0\trans\x_i)}
{(1-p_i)f(\bb_0\trans\x_i)} +o_p(1)\\
&=&n^{-1/2}\sumi \left\{m_0(\bb_0\trans\x_i) -E(Y_0)\right\}\\
&&+n^{-1/2}\sumi E\left\{
\frac{(1- T_j)\{Y_{0j}-m_0(\bb_0\trans\x_i)\}
K_h(\bb_0\trans\X_j-\bb_0\trans\x_i)}
{(1-p_i)f(\bb_0\trans\x_i)}\right\}\\
&&+n^{-1/2}\sumj E\left\{
\frac{(1- t_j)\{y_{0j}-m_0(\bb_0\trans\X_i)\}
K_h(\bb_0\trans\x_j-\bb_0\trans\X_i)}
{(1-P_i)f(\bb_0\trans\X_i)} \right\}\\
&&-n^{1/2} E\left\{
\frac{(1-T_j)\{Y_{0j}-m_0(\bb_0\trans\X_i)\}
K_h(\bb_0\trans\X_j-\bb_0\trans\X_i)}
{(1-P_i)f(\bb_0\trans\X_i)} \right\}+o_p(1)\\
&=&n^{-1/2}\sumi \left\{m_0(\bb_0\trans\x_i) -E(Y_0)\right\}\\
&&+n^{-1/2}\sumj 
E[1+\exp\{\eta(\ba\trans\X_j)\}\mid\bb_0\trans\x_j](1-t_j)\{y_{0j}-m_0(\bb_0\trans\x_j)\}
+o_p(1).
\ese
Combining the above results, we get
\bse
&&\sqrt{n}\{\wh E(Y_0)-E(Y_0)\}\\
&=&n^{-1/2}\sumi \left\{m_0(\bb_0\trans\x_i) -E(Y_0)\right\}\\
&&+n^{-1/2}\sumi
E[1+\exp\{\eta(\ba\trans\X_i)\}\mid\bb_0\trans\x_i](1-t_i)\{y_{0i}-m_0(\bb_0\trans\x_i)\}\\
&&
-n^{-1/2}\sumi E[\vec\{\X_{Li}\m_0'(\bb_0\trans\X_i)\trans\}]\trans
\B_0\\
&&\times (1-t_i)\{y_{0i}-m_0(\bb_0\trans\x_i)\}
\vec[\m_0'(\bb_0\trans\x_i) 
\otimes\{\x_{Li}-E(\X_{Li}\mid\bb_0\trans\x_i)\}] +o_p(1).
\ese

Now combining the results regarding $\wh E(Y_1)$ and $\wh E(Y_0)$, we
get
\bse
&&\sqrt{n}[\{\wh E(Y_1)-E(Y_1)\}-\{\wh E(Y_0)-E(Y_0)\}]\\
&=&n^{-1/2}\sumi \left\{
m_1(\bb_1\trans\x_i) 
- m_0(\bb_0\trans\x_i) -E(Y_1) +E(Y_0)\right\}\\
&&+n^{-1/2}\sumi 
E[1+\exp\{-\eta(\ba\trans\X_i)\}\mid\bb_1\trans\x_i]t_i\{y_{1i}-m_1(\bb_1\trans\x_i)\}\\
&&-n^{-1/2}\sumi
E[1+\exp\{\eta(\ba\trans\X_i)\}\mid\bb_0\trans\x_i](1-t_i)\{y_{0i}-m_0(\bb_0\trans\x_i)\}\\
&&-n^{-1/2}\sumi E[
\vec\{\X_{Li}\m_1'(\bb_1\trans\X_i)\trans\}]\trans
\B_1\\
&&\times t_i\{y_{1i}-m_1(\bb_1\trans\x_i)\}
\vec[\m_1'(\bb_1\trans\x_i) 
\otimes\{\x_{Li}-E(\X_{Li}\mid\bb_1\trans\x_i)\}]\\
&&
+n^{-1/2}\sumi E[
\vec\{\X_{Li}\m_0'(\bb_0\trans\X_i)\trans\}]\trans
\B_0\\
&&\times (1-t_i)\{y_{0i}-m_0(\bb_0\trans\x_i)\}
\vec[\m_0'(\bb_0\trans\x_i) 
\otimes\{\x_{Li}-E(\X_{Li}\mid\bb_0\trans\x_i)\}] +o_p(1).
\ese
\qed

\clearpage

\begin{table}[H]
\centering
\caption{Results for Study 1 based on 1000 replicates, where Full
  gives the average causal effect and corresponding standard
  deviation (sd) based on all potential responses, i.e. including the
  counterfactual ones not observable in practice, and Naive the same
  statistics based only on the observed potential responses. For the
  different estimators, we also compute the mean of the estimated sd
  (based on asymptotics, column $\wh{\rm sd}$), the empirical coverage
  obtained with confidence intervals based on these estimated sd (95\%
  cvg), and finally the mean squared error (mse).} 
 \label{tab:avgtreat}
 \vspace{0.2cm}
\begin{tabular}{c | c c c c c c c c}
Estimators & Full & Naive & IMP & IMP2 & IPW & AIPW & IAIPW & Shrinkage\\ \hline
mean & 2.030 & 1.569 &2.007 &2.032 &2.029 &2.037 &2.036 &2.036 \\ 
sd & 0.118 & 0.172 & 0.123 &0.122 &0.168 &0.131 &0.130 &0.131 \\
$\wh{\rm sd}$ & - & - & 0.134 &0.130 &0.176 &0.146 &0.146 &0.138  \\ 
95\% cvg & - & - & 96.1\% &96\% &96.5\% &97.8\% &98\% &97.5\% \\ 
mse & - & - & 0.016 &0.015 &0.028 &0.017 &0.017 &0.017 \\ \hline
\end{tabular}
\end{table}

\begin{table}[H]
\centering
\caption{Results for Study 2, where $m_1(\cdot)$ and $m_0(\cdot)$ are misspecified; see also caption of Table \ref{tab:avgtreat}.}
 \label{tab:avgtreatm}
 \vspace{0.2cm}
\begin{tabular}{c | c c c c c c c c}
Estimators& Full & Naive & IMP & IMP2 & IPW & AIPW & IAIPW & Shrinkage\\ \hline
mean & 3.990 & 3.647 &  3.761 &3.716 &4.005 &3.984 &3.979 &3.983 \\ 
sd & 0.137 & 0.202 &  0.187 &0.189 &0.207 &0.188 &0.189 &0.188 \\
$\wh{\rm sd}$ & - & - & 0.188 &  0.193 & 0.211 &  0.195 &  0.195 &  0.194 \\ 
95\% cvg & - & - &79\% &74.7\% &95.8\% &94.9\% &94.9\% &94.9\% \\ 
mse & - & - & 0.087 &0.111 &0.043 &0.035 &0.036 &0.035 \\ \hline
\end{tabular}
\end{table}

\begin{table}[H]
\centering
\caption{Results for Study 3, where $\eta(\cdot)$ is misspecified; see also caption of Table \ref{tab:avgtreat}.}
 \label{tab:avgtreatp}
 \vspace{0.2cm}
\begin{tabular}{c | c c c c c c c c}
Estimators  & Full & Naive & IMP & IMP2 & IPW & AIPW & IAIPW & Shrinkage\\ \hline
mean & 2.033 & 1.596 &2.009 &2.029 &2.030 &2.037 &2.037 &2.036 \\ 
sd & 0.122 & 0.165 & 0.123 &0.122 &0.169 &0.135 &0.134 &0.135 \\
$\wh{\rm sd}$ & - & - & 0.140 &0.140 &0.160 &0.143 &0.143 &0.142 \\ 
95\% cvg & - & - & 96.8\% &97.6\% &94.5\% &96\% &96.3\% &95.8\% \\ 
mse & - & - & 0.016 &0.015 &0.029 &0.018 &0.018 &0.018 \\ \hline
\end{tabular}
\end{table}

\begin{table}[H]
\centering
\caption{Results for Study 4, where $m_1(cdot)$, $m_0(\cdot)$, and $\eta(\cdot)$ are misspecified; see also caption of Table \ref{tab:avgtreat}.}
\label{tab:avgtreatmp}
\vspace{0.2cm}
\begin{tabular}{c | c c c c c c c c}
Estimators& Full & Naive & IMP & IMP2 & IPW & AIPW & IAIPW & Shrinkage\\ \hline
mean & 3.986 & 3.665 & 3.727 &3.637 &3.987 &3.980 &3.977 &3.980 \\ 
sd & 0.135 & 0.198 & 0.175 &0.173 &0.202 &0.186 &0.184 &0.186 \\
$\wh{\rm sd}$ & - & - & 0.194 & 0.205 & 0.207 & 0.191 & 0.191 & 0.191  \\ 
95\% cvg & - & - &78.5\%     &66.7\%    &95.4\%    &95.5\%     &96.1\%     &95.5\% \\ 
mse & - & - &0.098 &0.152 &0.041 &0.035 &0.034 &0.035 \\ \hline
\end{tabular}
\end{table}



\begin{table}[H]
\centering
\caption{Estimated average causal effect of maternal smoking on birth weight, including standard error and confidence interval, for the estimators introduced.}
\label{tab:realdata}
\vspace{0.2cm}
\begin{tabular}{c | c c c }
Estimator      &              Estimate      &      ${\rm se} $ &             $95$\% CI \\ \hline
naive           &             -275.3   &                   -        &      -           \\        
IMP              &            -259.8    &      22.2   &    (-303.3,-216.3) \\
IMP2           &      -262.6      &      23.1     &      (-307.8,-217.4)\\
IPW            &             -296.5   &       85.5    &      (-464.2,-128.9) \\
AIPW        &              -264.6     &      22.2      &      (-308.1,-221.1) \\
IAIPW        &              -264.7     &      22.2      &       (-308.3,-221.2) \\
Shrinkage    &           -264.6      &    22.2        &    (-308.1,-221.1) \\ \hline
\end{tabular}
\end{table}

\end{document}